\documentclass[pra,aps,twocolumn,superscriptaddress]{revtex4-2}
\usepackage{CJK}
\usepackage{graphicx}
\usepackage{dcolumn}
\usepackage{subfigure}
\usepackage{bm}
\usepackage{amsmath}
\usepackage{amssymb}
\usepackage{color,soul}
\usepackage{xcolor}
\usepackage{bbold}
\usepackage{mathptmx}
\usepackage[pagebackref=false,colorlinks,allcolors=blue!80,citecolor=blue]{hyperref}

\DeclareMathAlphabet{\mathcal}{OMS}{cmsy}{m}{n}

\newcommand{\ket}[1]{|#1\rangle}

\newcommand{\mean}[1]{\langle#1\rangle}

\newcommand{\half}{\tfrac{1}{2}}
\newcommand{\fsr}{\delta_\textsc{fsr}}

\begin{document}

\title{Accelerated Gaussian quantum state transfer between two remote mechanical resonators}

\author{M. Rezaei}
\affiliation{Department of Physics, Ferdowsi University of Mashhad, Mashhad, PO Box 91775-1436, Iran}

\author{K. Javidan}
\affiliation{Department of Physics, Ferdowsi University of Mashhad, Mashhad, PO Box 91775-1436, Iran}

\author{M. Abdi}
\email[Corresponding author; ]{mehabdi@gmail.com}
\affiliation{Department of Physics, Isfahan University of Technology, Isfahan 84156-83111, Iran}


\begin{abstract}
The main challenge in deterministic quantum state transfer between remote mechanical resonators is the local decoherence and the transmission losses in the communication channel.
In the path of overcoming this limitation, here we employ a shortcut to adiabatic passage protocol to devise a fast and reliable evolution path between two remote mechanical modes in separate optomechanical systems.
A quantum state transfer between the two nodes is conceived by engineering their coupling to an intermediate fiber optical channel.
The coupling pulses are operated such that the dark eigenmode of the system is decoupled from the fiber modes and transitions to the bright modes are compensated for by counterdiabatic drives.
We show that one obtains a quantum state transfer with high fidelity for various Gaussian states.
The efficiency is compared to that of adiabatic passage protocol in the presence of losses and noises.
Our results show that while the adiabatic passage protocol is very sensitive to the decoherence, the shortcut to adiabaticity provides a robust and fast quantum state transfer even for small values of the coupling strength.
The performance of both protocols are also investigated for the case of multimode fiber through numerical and an effective single-model model which is found by the elimination of off-resonant fiber modes.
Our findings may pave the way for using optomechanical systems in the realization of continuous-variable Gaussian quantum state transfer.
\end{abstract}
\maketitle

%
%
\section{Introduction}
A key task in quantum communications and quantum networks is the ability of quantum state transfer (QST) with high fidelity between two distant objects~\cite{kimble2008quantum, northup2014quantum, hammerer2010quantum}.
The quantum teleportation~\cite{Bennett1993, Braunstein1998} is a well-known method to accomplish this task, but today QST is a general terminology.
A quantum network consists of clusters of stationary quantum memories as nodes connected by quantum channels, such as free space or waveguides.
In general, through a QST protocol, the quantum information stored in the nodes are transferred from one to another via the channel~\cite{Bienfait2019, Kurpiers2018, Vermersch2017, ursin2007entanglement, ritter2012elementary}.
This task can essentially be performed probabilistically or deterministically.
In the former case a successful QST is conditioned on a post-selected entangled state between the node and is usually heralded upon a specific measurement outcome and is followed by a quantum teleportation at the end of the protocol~\cite{Olmschenk2009}.
Meanwhile, in the latter case the QST is always performed successfully but with a finite fidelity by mapping the state of the sending node onto an intermediator and then delivering it to the receiving node by properly engineering their mutual coupling~\cite{Axline2018}.

Among the others, there are two well-known protocols in performing deterministic QST between remote nodes of a quantum network.
The standard protocol which is based on wave packet shaping (WPS), employs laser drives for exciting the sending and receiving nodes so that the state is mapped into the flying qubits and subsequently perfectly absorbed at the destination~\cite{cirac1997quantum, Dantan2004}.
Another protocol for QST is based on the stimulated Raman adiabatic passage which has been developed to surpass the problems of WPS~\cite{Vogell2017}. During the adiabatic passage (AP) protocol, the quantum states are preserved in a dark state that decouples from the channel dissipation via driving pulses that are applied in a counterintuitive order~\cite{vitanov1997analytic}. 

In the realm of continuous variables~\cite{Braunstein2005}, optomechanical systems offer an excellent approach for implementing quantum communications with the mechanical resonators functioning either as a transducer or the stationary local nodes~\cite{stannigel2010optomechanical, McGee2013, Kurizki2015, Cernotik2016, Cernotik2019}. 
The QST between such mechanical nodes thus would be a crucial task.
As a preliminary step, the idea of transferring the state of light to a massive mechanical resonator had been proposed about two decades ago and later implemented experimentally~\cite{Parkins1999, Zhang2003, safavi2011proposal, Palomaki2013, Cernotik2018}.
Mechanical resonators in opto-electro-mechanical systems were later proposed to mediate the QST between electromagnetic modes in different frequencies~\cite{wang2012d, Barzanjeh2012, tian2012adiabatic, Schmidt2012, Sete2015, hill2012coherent, bochmann2013nanomechanical, andrews2014bidirectional, balram2016coherent}.
As the first step in networking with mechanical resonators, the QST between two mechanical modes and their entanglement has been the subsequent subject of research~\cite{habraken2012continuous, Xuereb2012, Abdi2012, Singh2012, shkarin2014optically, Abdi2015, Weaver2017, Zhang2020, Abdi2021}.
Specifically, the QST between remote mechanical modes through quantum teleportation has been theoretically investigated both in continuous variable~\cite{felicetti2017quantum} and discrete variable schemes~\cite{Pautrel2020, Li2020}.
However, creating highly entangled remote mechanical modes is a challenging task.
Therefore, a reliable and yet fast continuous variable QST protocol remains a rather unexplored area for mechanical networks.

Here, we propose and investigate an efficient and fast method for performing the QST between two remote mechanical modes without entangling them.
In our scheme each mechanical resonator is part of an optomechanical system that their cavity modes are connected via an optical fiber.
The mechanical modes are coupled to each other through the cavity and fiber modes thanks to the optomechanical interactions.
The cavity modes are separately driven by laser pulses with a proper detuning.
The shape of drive pulses are engineered such that the quantum state of one mechanical mode, the sending node, is transferred to the other one, the receiving node, without exciting the fiber modes.
This basically constitutes the adiabatic passage protocol where the system is preserved in a dark mode with respect to the mediator, here the fiber, during the transfer.
However, to speed-up the process and evade the destroying effect of local decoherence, we propose to compensate for the diabatic transitions via shortcut to adiabaticity~\cite{berry2009transitionless, chen2010shortcut}.
In this paper, we thoroughly investigate the shortcut to adiabatic passage (SAP) protocol for different system parameters in the presence of environmental effects.
In the SAP protocol, according to the transitionless quantum driving algorithm, the diabatic transitions among the adiabatic eigenmodes are suppressed by adding auxiliary counter-diabatic processes.
This leads to a fast and high fidelity state transfer through the dark mode evolution.
In contrast to the AP protocol which has a challenge in conflicting between transfer speed and efficiency, high fidelity QST becomes possible even for short operation times and even with small values of the coupling strength in the SAP protocol.

These protocols are applicable for transferring any quantum state between the mechanical resonators.
Nonetheless, here we put our focus on the Gaussian states.
These states are of great interest for their theoretical and experimental feasibility and yet their wide variety of applications~\cite{Weedbrook2012}.
Therefore, the performance of the AP and SAP protocols are compared for a range of Gaussian states that are of interest for continuous variable quantum information processing. 
We investigate the QST of coherent, squeezed vacuum, and squeezed coherent states with various amplitudes and squeezing parameters.

Furthermore, in our study we consider both the single- and multi-mode fiber cases.
For the latter, alongside a full numerical analysis an effective single-fiber mode model is analytically derived by adiabatic elimination of off-resonance fiber modes.
The validity of the effective mode is verified by its comparison to the numerical results.
It is worth mentioning that the other two major QST protocols, namely quantum teleportation and wave packet shaping, are not appropriate for the scheme studied in this work.
The former requires highly entangled mechanical resonators which is not easy to establish between remote sites as in our proposed setup.
Meanwhile, the latter is based on populating the optical fiber modes. Therefore, the fiber loss results-in extra loss and reduction in the QST fidelity.

The paper is organized as follows:
In  Sec.~\ref{sec:model} the setup model and its Hamiltonian is introduced.
In Sec.~\ref{sec:dynamics} the system dynamics is described through quantum Langevin equations and the cavity modes are adiabatically eliminated. Then the formulation of system dynamics and fidelity for the Gaussian states is discussed.
Sec.~\ref{sec:protocols} is devoted to describing the the AP and SAP protocols.
In Sec. \ref{sec:results} the numerical results of the QST through both protocols in the single- and multi-mode fiber cases are presented and discussed for various situations and states.
The work is summarized in Sec. \ref{sec:conclusion}.

%
%
\section{The model}\label{sec:model}
\begin{figure}[tb]
\includegraphics[width=\columnwidth]{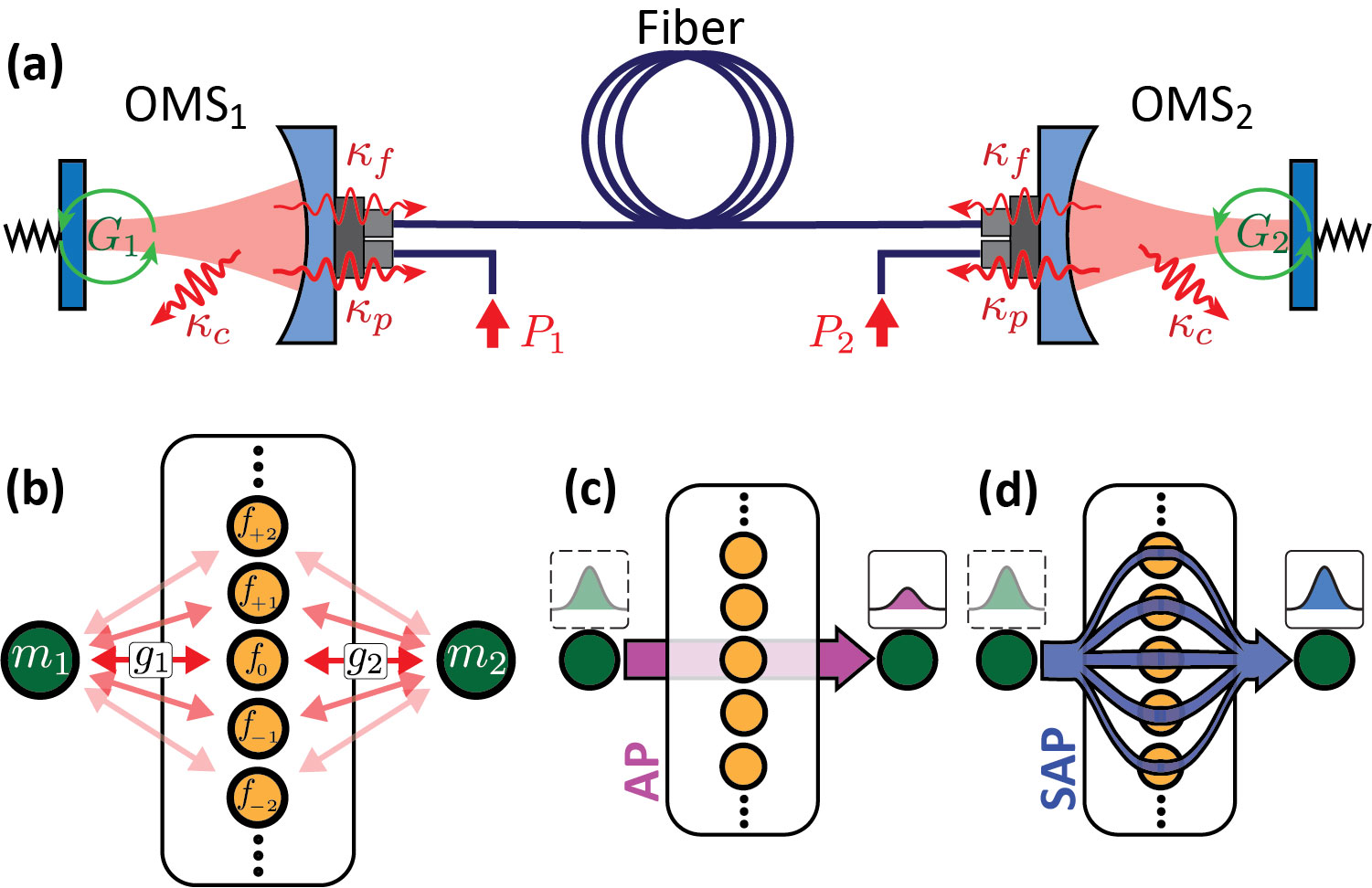}
\caption{%
(a) Sketch of the basic quantum network studied in this work: The nodes are optomechanical systems that are driven by their respective laser pulses ($P_i$ with $i=1,2$) and are connected to each other via an optical fiber.
(b) The simplified scheme: The mechanical resonators $m_1$ and $m_2$ are coupled to the fiber modes $f_{n}$ with effective strengths $g_1$ and $g_2$, respectively.
Schematic presentations for the adiabatic passage (c) and shortcut to adiabaticity (d) protocols; in AP the fiber modes are not excited, while in SAP the fiber excitations are retrieved and directed toward the receiving node during the process by compensating for the diabatic transitions.%
}%
\label{fig:scheme}%
\end{figure}

The system under study is composed of two similar nodes each containing an optomechanical system (OMS) whose cavity modes are connected via an optical fiber as shown in Fig.~\ref{fig:scheme}(a).
The cavity modes intermediate the interaction of mechanical resonators with the fiber modes. 

\subsection{The optomechanical system}
The OMS at each node is composed of a mechanical resonator and a cavity which is driven by a laser. The two interact with each other via radiation pressure~\cite{aspelmeyer2014cavity}. The Hamiltonian reads ($\hbar=1$)
\begin{equation}
H_{\rm om} = \omega_c a^\dag a +\omega_m m^\dag m + G_0 a^\dag a(m +m^\dag) +H_{\rm drv},
\label{OMS}
\end{equation}
where $G_0$ is the `bare' optomechanical coupling rate, while $\omega_c$ and $\omega_m$ are the bosonic cavity and mechanical mode frequencies, respectively.
Here, $a$ ($a^\dag$) and $m$ ($m^\dag$) respectively are the cavity and mechanical mode annihilation (creation) operators with the commutation relations $[a,a^\dag]=[m,m^\dag]=1$.
The system is driven by a laser at frequency $\omega_l$ whose corresponding Hamiltonian is given by
\begin{equation}
H_{\rm drv} = i\sqrt{\frac{\kappa_p P(t)}{\hbar\omega_l}}\ a^\dag e^{-i\omega_l t} +\text{H.c.},
\label{hdrv}
\end{equation}
where $P(t)$ is the input laser power and $\kappa_p$ is the rate of cavity decay into the pumping port.
The cavity photons can also decay into the fiber at the rate $\kappa_f$ or they are absorbed or diffracted inside the cavity as a consequence of the intrinsic loss effects, which we describe by the rate $\kappa_c$, see Fig.~\ref{fig:scheme}(a).
Note that in the Hamiltonian \eqref{hdrv} we are explicitly assuming a time dependent input power because in the following a pulsed scheme is studied.
Since the bare coupling rate $G_0$ is typically small one employs high drive powers to compensate for it.
As a consequence, the nonlinear nature of the optomechanical interaction becomes negligible.
An effective bilinear optomechanical Hamiltonian is then attained by substituting $a \to \mean{a} +c$ in Eq.~\eqref{OMS}, neglecting the term proportional to $c^\dag c(m +m^\dag)$, and dropping the drive.
Here, $c$ describes quantum fluctuations of the cavity filed around its mean classical value $\mean{a}$.
In the frame rotating at the laser frequency this effective OMS Hamiltonian reads~\cite{aspelmeyer2014cavity}
\begin{equation}
\widetilde{H}_{\rm om} = \Delta_c c^\dag c + \omega_m m^\dag m + G(t)(c +c^\dag)(m +m^\dag),
\label{omseff}
\end{equation}
where $\Delta_c = \omega_c-\omega_l$ is the cavity mode detuning from the laser frequency, while $G(t) =G_0\mean{a(t)}$ is the effective optomechanical coupling with $\mean{a}\propto \sqrt{P}$ the classical amplitude of the cavity mode~\cite{Abdi2011}.
Therefore, the effective optomechanical coupling becomes time dependent through the laser drive power.
The bilinear interaction in the above Hamiltonian involves two types of interactions: the beam-splitter $c^\dag m +c m^\dag$ and the two-mode squeezing $c m + c^\dag m^\dag$.
The former is suitable for transferring state of the mechanical mode to the cavity and vice versa~\cite{Palomaki2013}, while the latter can create entangled optomechanical states~\cite{Palomaki2013a}.
In the resolved sideband regime $\kappa \ll \omega_m$, where $\kappa\equiv \kappa_c+\kappa_p+\kappa_f$ is the total cavity decay rate, the dominant process is determined by setting the laser detuning: $\Delta_c = +\omega_m$ for the beam-splitter and $\Delta_c = -\omega_m$ for the two-mode squeezing.

\subsection{The fiber modes}
The fiber modes are treated as harmonic oscillators. The Hamiltonian is then given by
\begin{equation}
H_{\rm fib} = \sum_{n=-\infty}^{+\infty}\omega_{n} f_{n}^{\dagger}f_{n}=\sum_{n= -\infty}^{+\infty}(\omega_{0}+n\fsr) f_{n}^{\dagger }f_{n},
\end{equation}
where $\omega _{n} =\omega_{0} +n\fsr$ and $f_{n}$ ($f_{n}^\dag$) are the frequency and the annihilation (creation) operator of $n$th mode of the fiber, respectively.
Here, $\omega_{0}$ is frequency of the `resonant' fiber mode $f_0$, the one that has closest frequency to the laser drive $\omega_l$.
In the rightmost part of the above equation we have expressed the fiber mode frequencies with their separation from the `resonant mode' in terms of the free spectral range $\fsr=\pi c /L$ (with $c$ speed of light in the fiber and $L$ length of the fiber). 
In a frame rotating with the driving laser frequency one has
\begin{equation}
\widetilde{H}_{\rm fib} = \sum_{n= -\infty}^{+\infty}(\Delta_{0}+n\fsr) f_{n}^{\dagger }f_{n},
\end{equation}
where $\Delta_0 = \omega_0-\omega_{l}$ is detuning of laser from the \textit{resonant} fiber mode.

The fiber modes are coupled to the cavities at the nodes.
The cavity-fiber interaction is given by the following Hamiltonian~\cite{Vogell2017}
\begin{equation}
H_{\rm int} = i\chi\sum_{n=-\infty}^{+\infty}[c_1+(-1)^nc_2]f_n^\dag +\text{H.c.},
\label{fibcav}
\end{equation}
where $\chi=\sqrt{c\kappa_f/L} =\sqrt{\kappa_f\fsr/\pi}$ is the cavity-fiber interaction strength.
The odd fiber modes couple to cavity modes in the sending node with a relative phase difference of $\pi$ because of their odd number of maxima in the intensity profile.
The phase factor $(-1)^{n}$ in the Hamiltonian describes such phase differences.
Here, we have assumed that the \textit{resonant} mode is even.

Furthermore, we assume that the optomechanical systems at both sites are identical and for the sake of simplicity we set $\omega_0=\omega_c$.
Therefore, the Hamiltonian in Eq.~\eqref{fibcav} remains intact by moving to the laser frequency reference frame $\widetilde{H}_{\rm int} = H_{\rm int}$.

%
%
\section{System dynamics}\label{sec:dynamics}
The full system Hamiltonian in the laser rotating frame is given by
\begin{equation}
\widetilde{H} = \widetilde{H}_{\rm OM} +\widetilde{H}_{\rm fib} +\widetilde{H}_{\rm int},
\label{fullham}
\end{equation}
where the first term refers to the OMS at both sites:
\begin{equation}
\widetilde{H}_{\rm OM} = \sum_{i=1,2}\hspace{-1mm}\Delta_c c_i^\dag c_i + \omega_m m_i^\dag m_i + G_i(t)(c_i +c_i^\dag)(m_i +m_i^\dag).
\end{equation}

%
\subsection{Langevin equations}
The Langevin equations that give the system dynamics are found by
\begin{subequations}
\begin{align}
\dot{m}_i &= -i[m_i,\widetilde{H}] -\half\gamma_m m_i +\sqrt{\gamma_m}\ m_i^{\rm in}, \\
\dot{c}_i &= -i[c_i,\widetilde{H}] -\kappa_0 c_i +\sqrt{2\kappa_0}\ c_i^{\rm in}, \\
\dot{f}_n &= -i[f_n,\widetilde{H}] -\gamma_f f_n +\sqrt{2\gamma_f}\ f_n^{\rm in},
\end{align}
\label{langevin}%
\end{subequations}%
where $\kappa_0\equiv \kappa_c+\kappa_p$ accounts for intrinsic loss and channels of decay of the cavity other than the fiber, while $\gamma_m$ and $\gamma_f$ account for the mechanical damping and fiber loss rates, respectively.
Here, $m_i^{\rm in}$, $c_i^{\rm in}$, and $f_n^{\rm in}$ are the zero-mean Gaussian input noise operators with the following nonzero correlation functions
\begin{subequations}
\begin{align}
 \mean{m_i^{\rm in,\dag}(t),m_j^{\rm in}(t')} &= n_{\rm th}\delta_{ij}\delta(t-t'), \\
 \mean{m_i^{\rm in}(t),m_j^{\rm in,\dag}(t')} &= (n_{\rm th} +1)\delta_{ij}\delta(t-t'), \\
 \mean{c_i^{\rm in}(t),c_j^{\rm in,\dag}(t')} &= \delta_{ij}\delta(t-t'), \\
 \mean{f_n^{\rm in}(t),f_{n'}^{\rm in,\dag}(t')} &= \delta_{nn'}\delta(t-t'),
\end{align}%
\label{correlators}%
\end{subequations}%
where $n_{\rm th}$ is the thermal occupation number of the mechanical modes.
All other correlators are vanishing.

In the interaction picture of $H_{0}  = \sum_i (\omega _{m}m_i^\dag m_i +\Delta_c c_i^\dag c_i) +\sum_n \Delta_0 f_n^\dag f_n$, the equations in \eqref{langevin} read
\begin{align}
\dot{m}_i =& -\tfrac{\gamma_m}{2}m_i -iG_i\big[ c_i e^{-i(\Delta_c -\omega_m)t} +c_i^\dag e^{i(\Delta_c +\omega_m)t} \big] +\sqrt{\gamma_m} m_i^{\rm in}, \nonumber\\
\dot{c}_i =& -\kappa_0 c_i -i G_i\big[m_i e^{-i(\omega_m -\Delta_c)t} +m_i^\dag e^{i(\omega_m +\Delta_c)t} \big]\nonumber\\
&-\chi\sum_n (-1)^{n(i-1)} f_n e^{-i(\Delta_0 -\Delta_c)t} +\sqrt{2\kappa_0}c_i^{\rm in}, \\
\dot{f}_n =& -(\gamma_f +in\fsr)f_n +\chi[c_1 +(-1)^n c_2]e^{i(\Delta_0 -\Delta_c)t} \!+\!\sqrt{2\gamma_f}f_n^{\rm in},\nonumber
\end{align}
where in the second equation we have included the factor $(-1)^{n(i-1)}$ with $i=1,2$ for taking into account the phase difference in coupling of the cavity mode $c_2$ to the even and odd fiber modes.
In the two-mode squeezing regime ($\Delta_c=+\omega_m$) one could entangle the mechanical modes, which is useful for QST via teleportation. This regime is extensively studied in Ref.~\cite{felicetti2017quantum} and it is straightforward to apply their results to our proposed scheme.
Our analyses suggest that creating high entanglement between the mechanical resonators, which is necessary for a high fidelity continuous variable QST, is very demanding in the scheme studied in this paper.
Instead, we are interested in the beam-splitter regime which is the relevant case for the protocols studied in this work.
Therefore, we set $\Delta_c=-\omega_m$ and by assuming $G_i \ll \omega_m$ neglect the counter-rotating terms to arrive at
\begin{subequations}
\begin{align}
\dot{m}_i =& -\frac{\gamma_m}{2}m_i -iG_i c_i +\sqrt{\gamma_m} m_i^{\rm in}, \label{meclangevin}\\
\dot{c}_i =& -\kappa_0 c_i -i G_i m_i -\chi\sum_n (-1)^{n(i-1)} f_n +\sqrt{2\kappa_0}c_i^{\rm in}, \label{cavlangevin}\\
\dot{f}_n =& -(\gamma_f +in\fsr)f_n +\chi[c_1 +(-1)^n c_2]+\sqrt{2\gamma_f}f_n^{\rm in}. \label{fiblangevin}
\end{align}
\label{langevins}%
\end{subequations}%
We remind that since we have assumed $\omega_0 = \omega_c$, hence, one has $\Delta_0=\Delta_c$.
These equations can, in principle, be solve for the system dynamics.
Nonetheless, in the following we simplify the problem by adiabatically eliminating the cavity modes.

%
\subsection{Elimination of the cavity modes \label{ssec:adel}}
The cavity modes are employed to mediate the interaction of mechanical resonators (the nodes) to the fiber modes (the channel).
Assuming that there is a time-scale separation between the cavity dynamics and that of the fiber and mechanical modes, we adiabatically eliminate the cavity modes to attain an effective mechanical-fiber dynamics.
The cavity modes have larger decay rate compared to the fiber loss and the mechanical mode damping rates; $\kappa_0 \gg \gamma_m, \gamma_f$.
Therefore, if the decay rate is also larger than the coupling strengths of cavity to the fiber and mechanical modes ($\kappa_0 \gg G_i, \chi$) the time-scale separation in the dynamics is justified.

If the above mentioned condition is satisfied, the cavity mode reaches its steady-state much faster than the fiber and mechanical modes.
Hence, one sets the left hand side in \eqref{cavlangevin} equal to zero and finds the following equation for the steady-state cavity mode operators
\begin{equation}
c_i \approx -i\frac{G_i}{\kappa_0}m_i -\frac{\chi}{\kappa_0}\sum_n (-1)^{n(i-1)}f_n +\sqrt{\frac{2}{\kappa_0}} c_i^{\rm in},\hspace{1mm}(i=1,2)
\end{equation}
By plugging these back in \eqref{meclangevin} and \eqref{fiblangevin} one arrives at the following effective equations for the fiber and mechanical modes
\begin{subequations}
\begin{align}
\dot{m}_i =& -(\frac{\gamma_m}{2} +\frac{G_i^2}{\kappa_0})m_i -i \frac{\chi}{\kappa_0}G_i \sum_n (-1)^{n(i-1)}f_n +\xi_i, \\
\dot{f}_n =& -(\gamma_f +in\fsr)f_n -\frac{\chi^2}{\kappa_0} \sum_{k\in\mathbb{Z}}f_{n+2k} \nonumber\\
&-i\frac{\chi}{\kappa_0}[G_1 m_1 +(-1)^n G_2m_2] +\varphi_n,
\end{align}
\end{subequations}
where $G_i^2/\kappa_0$ is the damping rate induced in the mechanical through the optical cavity~\cite{patel2018single,stannigel2011optomechanical}.
Here, we have introduced the following noise operators
\begin{subequations}
\begin{align}
\xi_i \equiv& \sqrt{\gamma_m} m_i^{\rm in} -i\sqrt{2G_i^2/\kappa_0}\, c_i^{\rm in},
\label{mnoise}\\
\varphi_n \equiv& \sqrt{2\gamma_f} f_n^{\rm in} +\sqrt{2\chi^2/\kappa_0}\, [c_1^{\rm in} +(-1)^n c_2^{\rm in}].
\label{fnoise}
\end{align}%
\label{noises}
\end{subequations}
We notice that the cavity modes induce a coupling among the even and odd fiber modes, respectively.
This bilinear interaction is eliminated by applying a unitary transformation, see e.g.~\cite{Derezinski2017}.
In other words, one diagonalizes the fiber modes subspace through $f_n\to F_n=\sum_{n'}U_{n,n'}^{-1}f_{n'}$ with the transformation matrix $U$, arriving at
\begin{subequations}
\begin{align*}
\dot{m}_i =& -(\frac{\gamma_m}{2} +\frac{G_i^2}{\kappa_0})m_i -i \frac{\chi}{\kappa_0}G_i \sum_{n,n'} (-1)^{n(i-1)}U_{n,n'}F_{n'} +\xi_i,\\
\dot{F}_n =& -(\Gamma_{n} +i\Omega_n)F_n -i\frac{\chi}{\kappa_0}\sum_i\sum_{n'}(-1)^{n'(i-1)}U_{n,n'}^{-1}G_i m_i +\Phi_n,
\end{align*}
\end{subequations}
where the fiber normal mode loss rates $\Gamma_n$ and frequencies $\Omega_n$ are introduced and the normal fiber mode noise operators read $\Phi_n\equiv \sum_{n'} U_{n,n'}^{-1}\varphi_{n'}$.
In the working regime of our interest where $\chi \ll \kappa_0$, we find that the cavity induced coupling among the fiber modes is negligible compared to their frequency differences ($n\fsr\kappa_0 \gg \chi^2$).
Hence, we set $U\approx I$, where $I$ is the identity matrix.
Therefore, the normal fiber modes are the same as the `bare' modes $F_n\approx f_n$, as well as their resonance frequencies $\Omega_n \approx n\fsr$ and the corresponding noise operators $\Phi_n \approx \varphi_n$.
This is also numerically approved by diagonalizing the fiber subspace.
Nonetheless, the loss rates are modified as $\Gamma_n = \gamma_f +\chi^2/\kappa_0 \equiv \widetilde\gamma_f$.
By applying this approximation to the above equations we get
\begin{subequations}
\begin{align}
\label{efflangevina}%
\dot{m}_i =& -\widetilde\gamma_i\, m_i -i \sum_{n} g_{i,n}f_{n} +\xi_i, \\
\dot{f}_n =& -(\widetilde\gamma_f +in\fsr)f_n -i\sum_ig_{i,n} m_i +\varphi_n,
\label{efflangevinb}%
\end{align}%
\label{efflangevin}%
\end{subequations}
where we have introduced the total mechanical damping rates $\widetilde\gamma_i \equiv \gamma_m/2 +G_i^2/\kappa_0$ and the effective fiber-mechanical coupling strengths as
\begin{equation}
g_{i,n} =g_{i,n}(t) \approx (-1)^{n(i-1)}\frac{\chi G_i(t)}{\kappa_0}.
\label{coupling}
\end{equation}
In Fig.~\ref{fig:scheme}(b) a schematic for this effective model is given.

%
\subsection{Dynamics of Gaussian states\label{ssec:gaussian}}
The effective Langevin Eqs.~\eqref{efflangevin} are linear and the noise processes are Markovian with zero-mean Gaussian correlation functions.
In spite of that, analyzing the time evolution of an arbitrary state by their solution is cumbersome.
However, these linear equations are well-suited for describing dynamics of a system in which all mechanical and fiber modes are initially in a Gaussian state.
The Gaussian states span a wide range of states that are of interest for the ease of theoretical analysis and the experimental feasibility of their preparation, manipulation, and detection~\cite{Weedbrook2012}.
In terms of the Hermitian canonical operators $q_{1(2)}$ and $p_{1(2)}$---that are related to the creation and annihilation operators through $m_{1(2)}=(q_{1(2)} +ip_{1(2)})/\sqrt{2}$ and its Hermitian conjugate---such states are fully characterized by their first moments vector $d_j = \mean{Q_j}$ and the symmetrized covariance matrix $V_{jk}=\half\mean{\{Q_j-d_j,Q_k-d_k\}}$, where $\{,\}$ is the anticommutator and $Q\equiv [q_1,p_1,\cdots,X_n,Y_n,\cdots,q_2,p_2]^\intercal$ is the vector of mechanical and fiber quadrature operators.
Here, $f_n\equiv(X_n +iY_n)/\sqrt{2}$ gives the quadrature operators of the $n$th fiber mode.
Hence, a perfect QST for a Gaussian state is performed, provided the vector of mean values $d_1$ as well as all elements of the covariance matrix $V_1$ of the sending mode at time $t=t_i$ are transferred to the receiving site at the end of protocol $t=t_f$.

Time evolution of the displacement vector and the covariance matrix is readily found from the Langevin equations in \eqref{efflangevin}.
Since the noise operators have zero-mean values one arrives at the following equations that fully describe the system dynamics
\begin{subequations}
\begin{align}
 \dot{d} &= \mathcal{M}d, \\
 \dot{V} &= \mathcal{M}V +V\mathcal{M}^\intercal +\mathcal{D},
\end{align}%
\label{cmdynamics}%
\end{subequations}%
where $\mathcal{M}$ is the drift matrix, which in terms of quadrature operators is given by
\begin{equation*}
 \mathcal{M} \!= \!\hspace{-1mm}\left[\hspace{-1mm}
 \begin{array}{cccccccc}
   -\widetilde\gamma_1 & 0 & \cdots & 0 & g_{1,n} & \cdots & 0 & 0 \\
   0 & -\widetilde\gamma_1 & \cdots & -g_{1,n} & 0 & \cdots & 0 & 0 \\
   \vdots & \vdots & \ddots & \vdots & \vdots & \vdots & \vdots & \vdots \\
   0 & g_{1,n} & \cdots & -\widetilde\gamma_f & n\delta_{\rm FSR} & \cdots & 0 & g_{2,n}\! \\
   -g_{1,n} & 0  & \cdots & -n\delta_{\rm FSR} & -\widetilde\gamma_f & \cdots & -g_{2,n} & 0 \\
   \vdots & \vdots & \ddots & \vdots & \vdots & \ddots & \vdots & \vdots \\
   0 & 0 & \cdots & 0 & g_{2,n}\! & \cdots & -\widetilde\gamma_2 & 0 \\
   0 & 0 & \cdots &  -g_{2,n} & 0 & \cdots & 0 & -\widetilde\gamma_2 \\
 \end{array}\hspace{-1.5mm}\right]\!.
\label{drift}
\end{equation*}
Meanwhile, the diffusion matrix, i.e. the matrix of noise correlators reads $\mathcal{D}=D\otimes I_2$ with $I_2$, the $2\times 2$ identity matrix and
\begin{equation*}
 D \!=\!\hspace{-1mm}\left[\hspace{-1mm}
 \begin{array}{ccccccc}
  (\gamma_m n_{\rm th}+\widetilde\gamma_1) &\hspace{-1.5mm} \cdots & 0 & 0 & 0 & \cdots &\hspace{-1.5mm} 0 \\
  \vdots &\hspace{-1.5mm} \ddots & \vdots & \vdots & \vdots & \vdots &\hspace{-1.5mm} \vdots\hspace{-1.5mm} \\
  0 &\hspace{-1.5mm} \cdots & \widetilde\gamma_f  & 0 & \frac{\chi^2}{\kappa_0}  & \cdots &\hspace{-1.5mm} 0 \\
  0 &\hspace{-1.5mm} \cdots & 0 & \widetilde\gamma_f  & 0 & \cdots &\hspace{-1.5mm} 0 \\
  0 &\hspace{-1.5mm} \cdots & \frac{\chi^2}{\kappa_0}  & 0 & \widetilde\gamma_f  & \cdots &\hspace{-1.5mm} 0 \\
  \vdots &\hspace{-1.5mm} \vdots & \vdots & \vdots & \ddots & \vdots &\hspace{-1.5mm} \vdots \\
  0 &\hspace{-1.5mm} \cdots & 0 & 0 & 0 & \cdots &\hspace{-1.5mm} (\gamma_m n_{\rm th}+\widetilde\gamma_2) \\
 \end{array}\hspace{-1.5mm}\right].
\label{diffusion}
\end{equation*}

%
\subsection{Fidelity of Gaussian states}
In order to quantify the QST efficiency we employ the Gaussian state quantum fidelity to compute the overlap of the initial mechanical state at the sending node $\rho_i = \rho_1(t=t_i)$ with the state of the mechanical mode at the receiving node at the end of process $\rho_f = \rho_2(t=t_f)$. This quantity is computed by
\begin{equation}\label{fidelity}
\mathcal{F}(\rho_i,\rho_f) = \mathcal{F}_0(V_i,V_f)\exp\Big\{\hspace{-1.5mm}-\frac{1}{4}\delta d^\intercal(V_i+V_f)^{-1}\delta d\Big\},
\end{equation}
with $\delta d = d_f-d_i$~\cite{Banchi2015}.
The first factor only depends on the covariance matrices and can be expressed in terms of the symplectic invariants $\Upsilon = \det(V_i+V_f)$ and $\Lambda = 4\det(V_i +i\varpi)\det(V_f +i\varpi)$. Here, the symplectic matrix is
\begin{equation}
\varpi = \frac{1}{2}\left(
\begin{array}{cc}
	0 & 1 \\
	-1 & 0 \\
\end{array}
\right).
\end{equation}
For the single-mode states that we are interested in this work one arrives at
\begin{equation}
\mathcal{F}_0^2(V_i,V_f)=\frac{1}{\sqrt{\Upsilon +\Lambda}-\sqrt{\Lambda}}.
\end{equation}
In the rest of this paper we use Eq.~\eqref{fidelity} for quantifying the efficiency of the state transfer protocols.
The studied states are coherent squeezed states $\ket{\alpha,r}$ where $\alpha$ is in general a complex number denoting displacement of the state in the phase space. The degree of squeezing of the phase space distribution is encoded in $r$, where without loss of generality is taken a real number.

%
%
\section{Transfer protocols}\label{sec:protocols}
We employ the simple single-mode case to provide an intuitive picture for the formulation of the protocols studied in this work.
The compact form of the Langevin equations in Eq.~\eqref{efflangevin} reads
\begin{equation} 
\dot{u}(t)=-iM(t)u(t)+n(t),
\label{heisenberg}
\end{equation}
where $u$ and $n$ are the vector of mode and noise operators, respectively, and $M$ is the dynamical matrix or matrix of coefficients.
For a single-mode fiber, where only the resonant fiber mode couples to the mechanical nodes, one gets $u = (m_1, f_0, m_2)^\intercal$, $n = (\xi_1, \varphi_0, \xi_2)^\intercal$, and
\begin{equation}\label{dynamics}
M =
\left(\begin{array}{ccc}
		-i\widetilde\gamma_1 & g_{1} & 0 \\
		g_{1} & -i\widetilde\gamma_f & g_{2} \\
		0 & g_{2} & -i\widetilde\gamma_2
\end{array}\right),
\end{equation}
where we have used $g_{i} = g_{i,0}$ for shorthand.

%
\subsection{Adiabatic passage}
To begin, we first describe the adiabatic passage (AP) which is a well-known universal protocol for deterministic QST in discrete variable systems~\cite{Vogell2017}.
In the AP protocol, the laser drives are applied at the nodes in a counterintuitive order, such that the receiving node excitation $g_{2}(t)$ precedes that of the sending node $g_{1}(t)$.
This requirement implies that $\lim_{t\to t_{i}}\frac{g_{1}}{g_{2}}= 0$ and  $\lim_{t\to t_{f}}\frac{g_{1}}{g_{2}}= \infty$, which by introducing the mixing angle $\vartheta \equiv \arctan(\frac{g_{1}}{g_{2}})$ translates into $\vartheta(t_{i})=0$ and $\vartheta(t_{f})=\frac{\pi}{2}$ \cite{vitanov1997analytic}.
We remind that according to Eq.~\eqref{coupling} the fiber-mechanical coupling rates are proportional to the linearized optomechanical coupling rates $g_i(t) \propto G_i(t)$.
Therefore, the desired timing and form of interaction pulses $g_i$ is basically controlled via the laser drives $P_i$.

In analogy to the dark and bright states in atom systems, in our simple three-mode model the system eigenmodes comprise the dark mode $A_{0}$ and two bright modes $A_{\pm}$ whose relation to the original modes in the case of vanishing loss and damping rates ($\widetilde\gamma_{1(2)}=\widetilde\gamma_f=0$) are found as the following
\begin{subequations}\label{eigmodes}
\begin{align}
A_{+} &=\frac{1}{\sqrt{2}}\big( m_{1}\sin{\vartheta} + m_{2}\cos{\vartheta}+f_{0} \big), \\
A_{0} &=m_{1} \cos{\vartheta} -m_{2} \sin{\vartheta}, \\
A_{-} &=\frac{1}{\sqrt{2}}( m_{1}\sin{\vartheta} + m_{2}\cos{\vartheta}-f_{0}).
\end{align}
\end{subequations}
These `adiabatic modes' are the instantaneous eigenmodes of the system.
The AP mechanism is then easily understood by looking at Eqs.~\eqref{eigmodes}:
The goal is to transfer a quantum state from sending mechanical mode $m_1$ to the receiving mechanical mode $m_2$ through the dark mode $A_{0}$ that does not involve the fiber mode $f_0$.
The AP protocol requires large operation times for ensuring a slow evolution and/or strong driving pulses to maintain the system in the dark mode $A_{0}$ during the process.
Generically, if the process duration or driving pulses strength is large enough, the system evolves along either of its adiabatic eigenmodes $A_{i}$ without any intermode transitions.
Therefore, slow nature of the AP protocol is its main weakness.
Because it then suffers from dissipations and decoherence in the sending and receiving nodes as well as the loss in the channel.
On the other hand, if the process is accelerated for minimizing the losses, due to the diabatic transitions the time evolution path does not follow the adiabatic eigenmode $A_{i}$ anymore.
Such transitions, in turn, can be compensated for by introducing extra drives to the system as is detailed next.

\subsection{Shortcut to adiabatic passage}
Here, we propose a mechanism based on the shortcut to the adiabatic passage (SAP) technique to speed up the process and yet retain the high QST efficiency.
In principle, one needs to revert the diabatic processes among the adiabatic eigenmodes during the transfer~\cite{GueryOdelin2019}.
For this, we modify the dynamic matrix $M(t)$ in Eq.~\eqref{heisenberg} such that the non-adiabatic transitions are eliminated~\cite{chen2010shortcut}.
This allows us to perform a rapid and high fidelity QST between the mechanical resonators in the system even with weak driving pulses.
This, indeed, is performed by adding an auxiliary counter-diabatic process to the system whose contribution to the dynamics is described by the following dynamical matrix
\begin{equation}\label{cd}
M_{\rm cd}(t)= \sum_{k}\dot{A}^{}_{k}A_{k}^{\dagger},
\end{equation}
where the summation is over all eigenmodes of the system and the dot indicates the time derivative.
By substituting Eqs.~\eqref{eigmodes} in \eqref{cd} the counter-diabatic dynamical matrix $M_{\rm cd}$ for the three-mode system reads
\begin{equation} \label{Mcd}
M_{\rm cd}= \left(\begin{array}{ccc}  0 & 0 & ig_{a} \\ 0 & 0 & 0 \\ -ig_{a} & 0 & 0  \end{array}\right),
\end{equation}
where $g_{a}=(\dot{g}_{1}g_{2}-g_{1}\dot{g}_{2})/(g_{1}^{2}+g_{2}^{2})$ is the coupling rate of the mechanical modes to each other, e.g. via an auxiliary driving field.
For its realization, one should employ an extra field that couples to both mechanical modes at the sending and receiving nodes, which can be experimentally challenging for two remote resonators.

We instead propose an experimentally feasible approach for implementing the counter-diabatic processes.
For this, we absorb $M_{\rm cd}(t)$ into the reference pulses and avoid the need for an additional long range coupling.
The system dynamics satisfies SU(3) Lie algebra, i.e., in the absence of damping and losses the total drift matrix reads $M_{\rm tot}\equiv M+M_{\rm cd}=g_1\mathcal{G}_1+g_2\mathcal{G}_6 -g_a\mathcal{G}_5$, where $\mathcal{G}_k$ with ($k=1,\cdots,8$) are the Gell-Mann matrices.
By applying the unitary transformation
\begin{equation} \label{eq+10}
U(t)= \left(
\begin{array}{ccc}
	{1} & {0} & {0} \\
	{0} & {\cos\phi(t)} & {i\sin\phi(t)} \\
	{0} & {i\sin\phi(t)} & {\cos\phi(t)}
\end{array}\right)
\end{equation}
on $M_{\rm tot}$ we arrive at
\begin{equation} \label{eq+11}
\widetilde{M}_{\rm tot}(t)= \tilde{g}_{1}\mathcal{G}_{1}+ \tilde{g}_{2}\mathcal{G}_{6}-\tilde{g}_{a}\mathcal{G}_{5},
\end{equation}
where the following parameters are introduced
\begin{subequations}\label{gellmann}
\begin{align}
\tilde{g}_{1} &={g}_{1}\cos\phi-{g}_{a}\sin\phi, \\
\tilde{g}_{2} &={g}_{2}+\dot\phi, \\
\tilde{g}_{a} &={g}_{1}\sin\phi+{g}_{a}\cos\phi.
\end{align}
\end{subequations}
By setting $\tilde{g}_{a}=0$ one finds $\phi=-\arctan({g}_{a}/{g}_{1})$ and
\begin{subequations}
\begin{align}
\tilde{g}_{1} &=\sqrt{{g}_{1}^{2}+{g}_{a}^{2}}, \\
\tilde{g}_{2} &={g}_{2}+\dot\phi.
\end{align}%
\label{effcoup}%
\end{subequations}%
Therefore, the SAP protocol can be realized simply by modifying the local coupling pulses to $\tilde{g}_{1}$ and $\tilde{g}_{2}$~\cite{Zhang2019}.

In the following, we perform a numerical analysis on the SAP protocol by properly modifying the drift $\mathcal{M}$ and diffusion $\mathcal{D}$ matrices in Eqs.~\eqref{cmdynamics} and the two following equations and show that it allows for a fast and high fidelity state transfer between two mechanical modes by only reshaping the original driving pulses.
Note that the driving pulses $\tilde{g}_1$ and $\tilde{g}_2$ no longer need to satisfy the adiabaticity conditions such as large duration time or strong amplitude.

%
%
\section{Results and discussion}\label{sec:results}
\begin{figure}[t]
\includegraphics[width=\columnwidth]{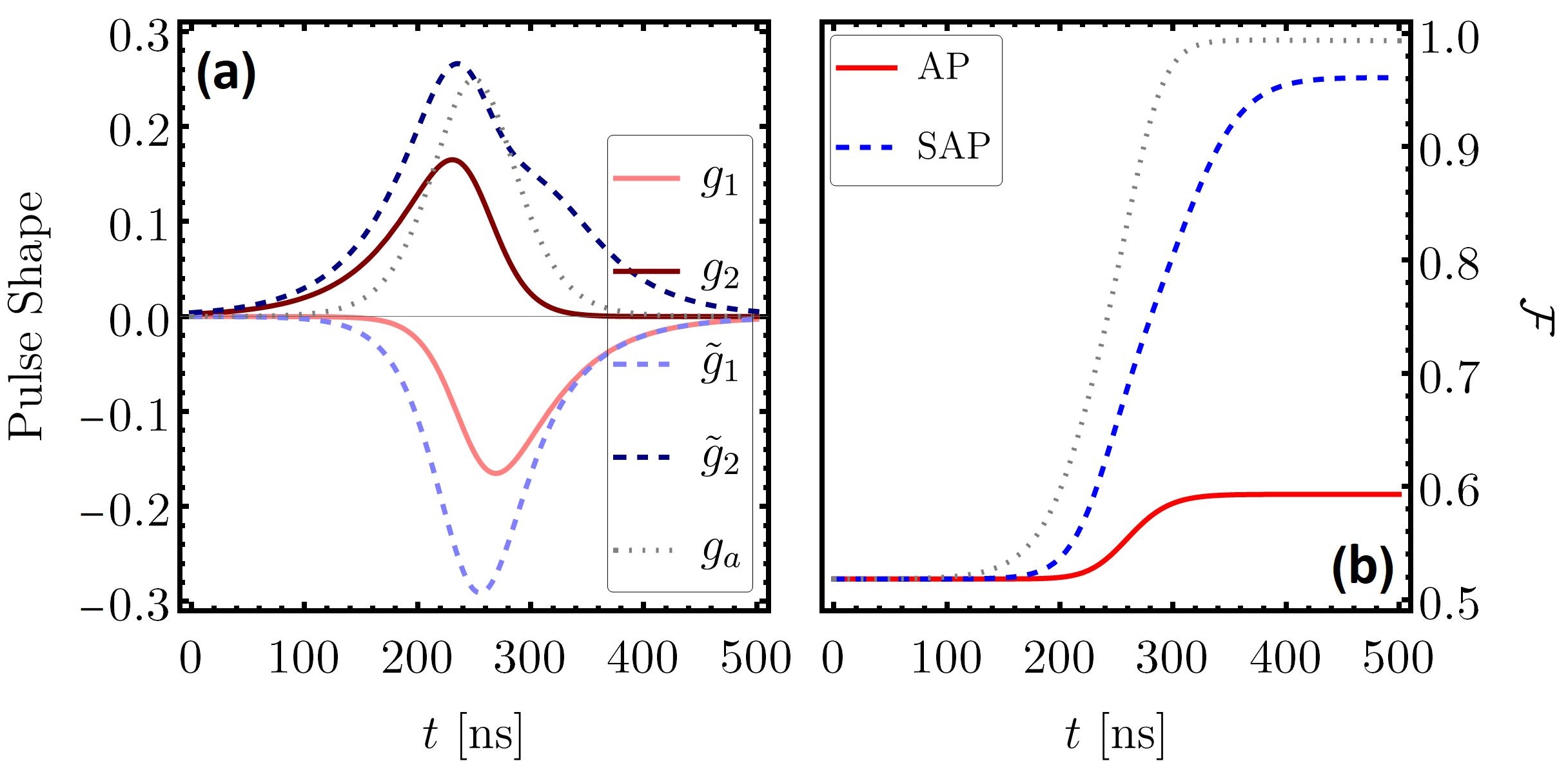}
\caption{%
(a) The original (solid) and modified (dashed) pulse shapes used for performing AP and SAP protocols. Also, the auxiliary pulse for performing a direct counter diabatic drive is shown as the dotted line. The vertical axis is normalized to $\lambda_0$ and the parameters are $T=500$~ns and $\sigma=0.1T$.
(b) Time evolution for the fidelity of the transferred coherent squeezed state $\ket{\alpha=1,r=1}$ by these two protocols.
The dotted line corresponds to the case where counter diabatic pulses applied directly through Eq.~\eqref{Mcd}. Here, we have used $\lambda_0/2\pi=10$~MHz, $\fsr = 5\kappa$, and $n_{\rm th}=0$.
}
\label{fig:pulse}
\end{figure}

We now study the QST by the shortcut to the adiabaticity method introduced and studied in the previous section and compare its efficiency to the adiabatic passage.
For performing the quantum state transfer, we choose the laser driving pulses such that the time-dependent optomechanical coupling rates read
\begin{subequations}\label{pulse}
\begin{align}
G_{1}(t) &=-\lambda_{0}\,\text{sech}[\frac{1}{\sigma}(t-\half T)] \sin[\frac{\pi}{4} s(t)], \\
G_{2}(t) &=+\lambda_{0}\,\text{sech}[\frac{1}{\sigma}(t-\half T)] \cos[\frac{\pi}{4} s(t)],
\end{align}
\end{subequations}
where $s(t)=1+\tanh[\frac{1}{\sigma}(t-\half T)]$ with $T$ the drive pulse duration and $\sigma$ its semi-width at the half-maximum.
The effective fiber-mechanical coupling strengths are then immediately found by Eq.~\eqref{coupling}.
Here, $\lambda_0$ is the maximum coupling in the pulse.
In Fig.~\ref{fig:pulse}(a) the typical form of the original pulses for an adiabatic state transfer and the modified pulse shapes suited for the SAP protocol are shown. In the following analysis we shall employ these pulses having $t_i=0$ and $t_f=T$. We also assume that the receiving mode is initially cooled to its ground state, which essentially can be performed by some mechanism, e.g., sideband cooling~\cite{aspelmeyer2014cavity}.
\begin{table}[b]
\caption{\label{tab:params}Parameters of the system.}
\begin{ruledtabular}
\begin{tabular}{lcrr}
\textrm{Parameter}&
\textrm{Symbol}&
\textrm{Value}&
\textrm{Unit}\\
\colrule
mechanical frequency & $\omega_{m}$ & $1(2\pi)$ & GHz\\
mechanical damping rate & $\gamma_{m}$ & $1(2\pi)$ & kHz\\
cavity frequency & $\omega_{c}$ & $193(2\pi)$ & THz \\
optomechanical coupling rate & $\lambda_0$ & $0-100(2\pi)$ & MHz \\
fiber loss rate & $\gamma_f$ & $1.5(2\pi)$ & kHz \\
branching ratio & $\eta$ & $0.025-0.557$ & \multicolumn{1}{c}{---} \\
\end{tabular}
\end{ruledtabular}
\end{table}

In the following sections we employ parameters from the typical optomechanical systems in the telecom wavelengths, i.e. $\omega_c/2\pi =193$~THz, where the commercial optical fibers have their lowest loss rates $\gamma_f/2\pi =1.5$~kHz.
Photonic crystal optomechanical systems and microresonators are known to have state of the art properties around the telecom frequencies~\cite{Park2009, Chan2011}.
The mechanical vibrations in such setups are high quality modes with frequencies in the range of a few hundreds of megahertz to a few gigahertz.
Hence, we consider $\omega_m/2\pi = 1$~GHz and $\gamma_m/2\pi = 1$~kHz in our numerics~\cite{Schliesser2009, Riedinger2016}.
The optomechanical coupling rate in such setups can reach to values as high as $\lambda_0 = \max\{G_{1(2)}(t)\} = 2\pi \times 50$~MHz, see e.g. Ref.~\cite{Ding2011, Krause2015}.
In this work we always set the cavity decay rate to $\kappa_0 = 5\lambda_0$.
Furthermore, the branching ratio is chosen such that one always has $\kappa_0=5\chi$.
These choices of the parameters guarantee the validity of the cavity adiabatic elimination performed in Sec.~\ref{ssec:adel} and ensure that the system operates at the resolved sideband regime.
The system parameters employed in our work are summarized in Table~\ref{tab:params}.

%
\subsection{Single-mode fiber}
First, we consider the simple case of a single-mode fiber and study various aspects of the QST protocols.
The single-mode fiber regime is guaranteed for short distances where $\fsr \gg \kappa$.
Therefore, we set $\fsr = 5\kappa$.
By employing the pulse shapes introduced above and shown in Fig.~\ref{fig:pulse}(a), we numerically solve for the dynamics of the system through Eqs.~\eqref{cmdynamics} and compute fidelity of the state at the receiving mechanical mode with respect to the initial state of the sending mechanical mode using Eq.~\eqref{fidelity}.
The instantaneous fidelity for transfer of a squeezed coherent state with $\ket{\alpha=1,r=1}$ as a representative Gaussian state under the AP and SAP protocols are shown in Fig.~\ref{fig:pulse}(b) for pulses with duration $T=500$~ns, width $\sigma =0.1T$, and strength $\lambda_0/2\pi = 10$~MHz.
For this coupling strength value a branching ratio of $\eta=0.025$ gives $\chi=5\kappa_0$ and $\fsr=5\kappa$ as required for the validity of our single-mode theory.
This also corresponds to the fiber length of $L\approx 0.58$~m.

\begin{figure}[tb]
\includegraphics[width=\columnwidth]{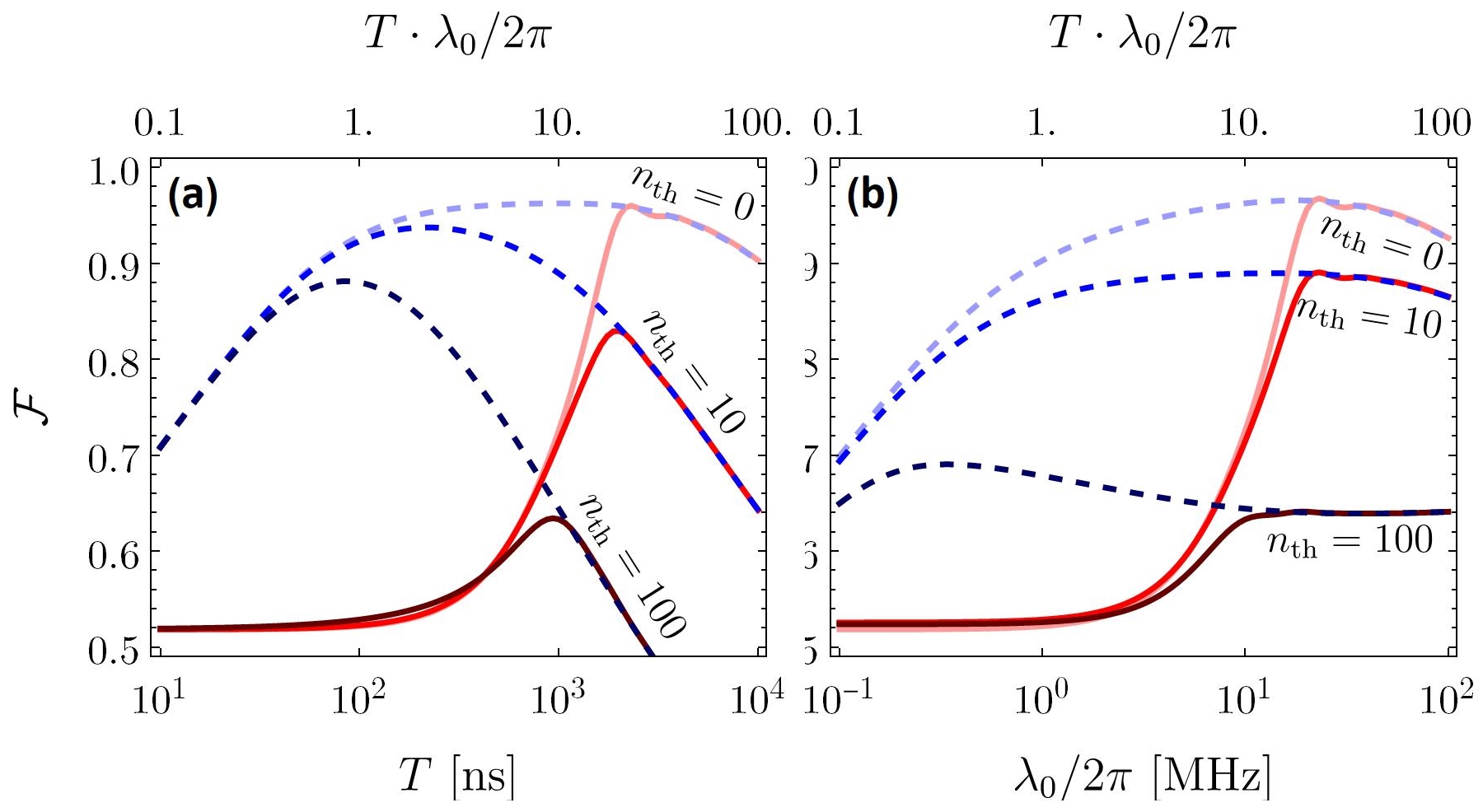}
\caption{The QST fidelity through a single-mode fiber:
(a) Variations with respect to the pulse duration for the coupling rate $\lambda_0/2\pi=10$~MHz and three different bath temperatures.
(b) As a function of pulse amplitude at the fixed $T=1000$~ns and various thermal noise amplitudes.
The solid and dashed lines are for the AP and SAP protocols, respectively.
}
\label{fig:single}
\end{figure}

The receiving mode is initially in a vacuum state. Hence, its overlap with the target state is nonvanishing ($\mathcal{F}\approx 0.52$).
Nonetheless, the final fidelity dramatically depends on the shape of the pulses: while the fidelity by SAP approaches $96\%$, the adiabatic passage only tops to about $60\%$ for the considered parameters.
It is also useful to check efficiency of the `original' SAP protocol where the counterdiabatic drives are applied independently as in Eq.~\eqref{Mcd}.
The dotted line in Fig.~\ref{fig:pulse}(b) shows the result and clearly indicates that it outperforms the modified SAP protocol.
This is mainly due to omitting the irreversible processes in obtaining the effective SAP coupling pulses $\tilde{g}_{1(2)}$.
However, as mentioned before, the original protocol requires mechanical-mechanical strong coupling that can be a challenging experimental task, specially for remote sites.
Therefore, from now on we only consider the modified SAP protocol where the pulses are only applied locally.

We now turn to study the effect of pulse duration and strength on the performance of the protocols. Here, and in the rest of the paper we set the pulse width to the tenth of pulse duration $\sigma=0.1T$ without loss of generality.
In Fig.~\ref{fig:single}(a) the variations of fidelity $\mathcal{F}$ in quantum transfer of the state $\ket{\alpha=1,r=1}$ from $m_1$ to the $m_2$ mode is plotted with respect to the pulse length for the two protocols.
We observe that a longer pulse does not guarantee a better outcome.
In fact, at longer times the decoherence at the nodes and in the channel degrades the transferred state.
Here, we have assumed different environmental temperatures at the nodes, whose effect appears as the equilibrium occupation numbers $n_{\rm th}=\{0,10,100\}$.
Evidently, the thermal noise increases the decoherence at the nodes, limiting the transfer fidelity.
We remind that the optical frequencies considered in this work are not appreciably affected by the temperature.
Fig.~\ref{fig:single}(a) shows that for both protocols a longer pulse duration only gives a better result if the temperature at the nodes is very low.
Yet, there is an optimal pulse duration at which the transfer process is fast enough to overcome the decoherence effects.

In Fig.~\ref{fig:single}(b) the role of coupling rate is investigated for $T=1000$~ns.
Since the mechanical damping rates and noises depend on $G_i$ through $\widetilde\gamma_i$, an enhanced pulse amplitude results-in an increase in the decoherences, too.
Hence, one observes a global maximum for the coupling rate amplitude $\lambda_0$ even at zero-temperature where the QST is performed optimally.
However, similar to the pulse duration $T$, the optimal value of $\lambda_0$ is different for AP and SAP protocols and also it depends on the temperature of the nodes.

From the curves in Fig.~\ref{fig:single} one observes that the general behavior of fidelity more or less depends on the product $\lambda_0T$.
In particular, the effect of thermal noise on the QST is not noticeable for small values of the product $\lambda_0T$.
The AP becomes sensitive to the thermal noise about $T\lambda_0/2\pi\approx 10$, while this number is about $\approx0.2$ for the SAP.
It is also concluded that in the AP protocol as the speed of the QST process increases the transfer efficiency decreases.
Because by speeding up transfer process in the AP protocol the evolution of system evolution does not follow the dark mode and the fiber mode can also get populated.
However, in the SAP protocol the diabatic processes are compensated for, thus it retains the performance even in the shorter pulse durations.

\begin{figure}[tb]
\includegraphics[width=\columnwidth]{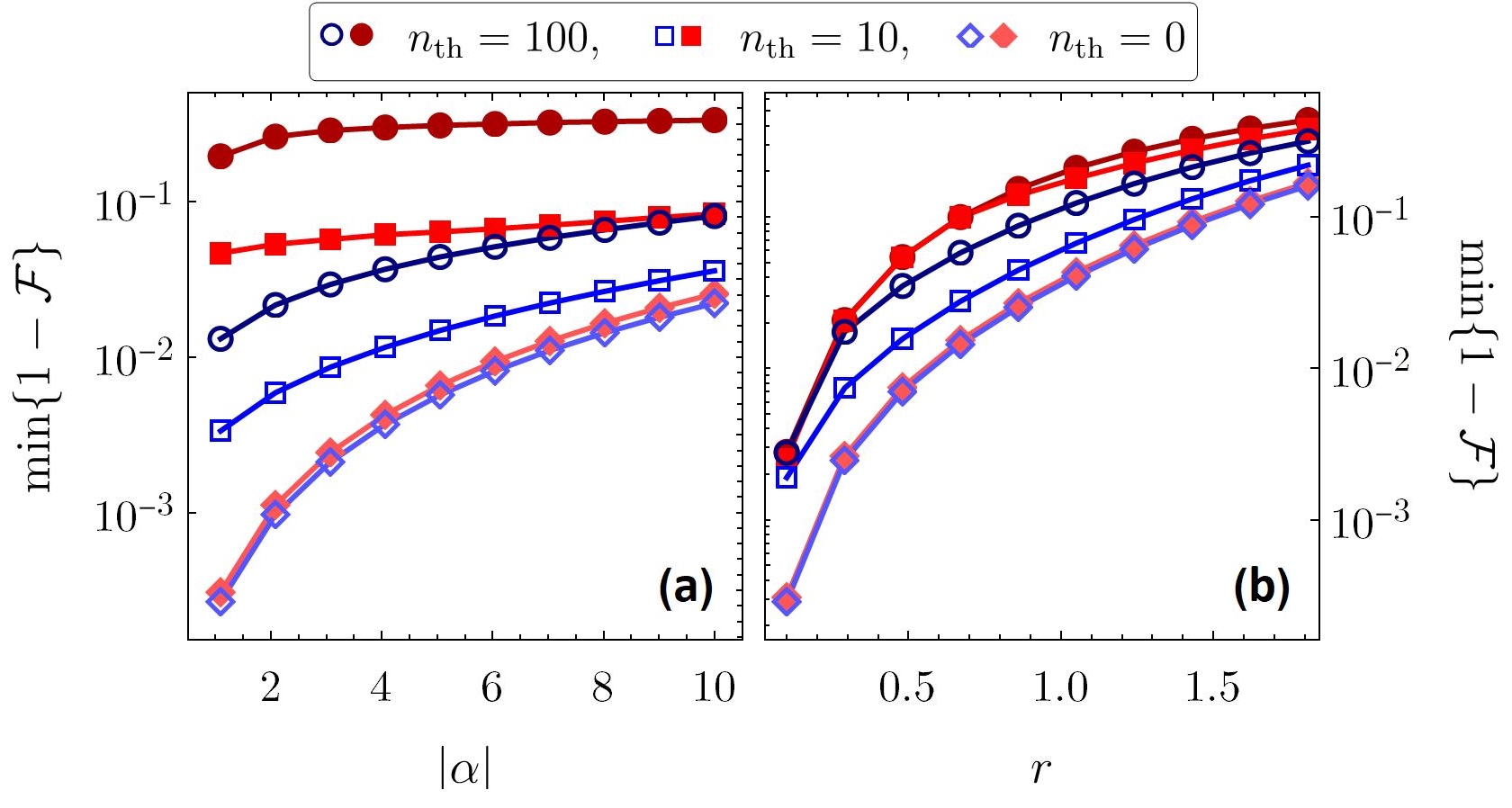}
\caption{The optimal quantum state transfer for
(a) a coherent state $\ket{\alpha,r=0}$ and
(b) a squeezed vacuum state $\ket{\alpha=0,r}$ via AP (filled markers) and SAP (open markers) protocols at three different temperatures.
Here, we have set $\lambda_0/2\pi=10$~MHz and the optimization is performed over $T$.
}
\label{fig:optimal}
\end{figure}

Finally, in Fig.~\ref{fig:optimal} we compare the optimal performance of the protocols for various coherent and squeezed vacuum states at three different temperatures expressed in terms of the mechanical occupation number $n_{\rm th}$. In producing these plots we have assumed a given coupling rate of $\lambda_0/2\pi = 10$~MHz and the optimization is only carried over the pulse duration.
The figure shows that the SAP protocol always outperforms that of adiabatic passage. This is more notable in the case of coherent states [Fig.~\ref{fig:optimal}(a)].
At very low ambient temperature $n_{\rm th} = 0$ both protocols result-in almost the same fidelities.
Nonetheless, as the temperature rises the SAP gives significantly better outcome for coherent states.
Furthermore, the optimal fidelity in QST of a coherent state becomes almost independent of its amplitude $|\alpha|$ at higher temperatures.
In contrast, for a squeezed vacuum state the performance of both protocols gets more limited at larger squeezing parameter values $r$ for all temperatures.
This is such that even for zero thermal noise $n_{\rm th}=0$ the error in QST of the state $\ket{\alpha=0,r=2}$ is about fifteen percent.
Interestingly, the curves in Fig.~\ref{fig:optimal}(b) suggest that for $n_{\rm th}\gtrsim 10$ the performance of the SAP protocol for weakly squeezed vacuum states ($r\lesssim 0.2$) is more or less the same and independent of temperature.
This number extends to $r\lesssim 0.9$ for the adiabatic passage.
We eventually must emphasize that our results clearly show that the efficiency of the protocols depends on the state being transferred.
This mainly stems from the fact that different states have different sensitivities to the noise and decohering effects.

%
\subsection{Multi-mode fiber}
We now turn to investigate the effect of higher order modes of the fiber on the transfer efficiency in the protocols.
To begin, we first restrict our calculations to the fiber central mode $f_{0}$ and the first pair of neighboring modes $f_{\pm 1}$.
This brings us to a set of Langevin equations of motion as in Eq.~\eqref{heisenberg} with the vector operator $u=(m_{1}, f_{-1}, f_{0}, f_{+1}, m_{2})^\intercal$. Since we are putting our focus on the transfer of quantum Gaussian states the dynamics of the first and second moments are then given by Eqs.~\eqref{cmdynamics} and the corresponding drift and diffusion matrices are found as outlined at the end of Sec.~\ref{ssec:gaussian}.
The eigenmodes of the system with three fiber modes in the absence of decoherences are straightforward to find. The `darkest' mode recalling the sign of interactions ($g_{1,0}=+g_{1,\pm1}\equiv g_1$ and $g_{2,0}=-g_{2,\pm1}\equiv g_2$) reads
\begin{equation}
A_{0} = \mathcal{N}_0
\Big({g_{2}}, {\frac{2g_{1}g_{2}}{\fsr}}, 0, {-\frac{2g_{1}g_{2}}{\fsr}}, {-g_{1}}\Big)^{\intercal},
\label{darkmode}
\end{equation}
where $\mathcal{N}_0$ is the normalization factor.
Even though the normal mode $A_{0}$ is dark with respect to the  fiber mode $f_{0}$, it still has a finite coupling to the first sideband modes $f_{\pm1}$ and hence these modes can get populated during the QST process.
In other word, the adiabatic mode $A_{0}$ is not totally dark when other modes of the fiber are taken into account.
Nonetheless, the contribution of the sideband modes is inversely proportional to the fiber free spectral range. Meaning that this adiabatic mode can still behave as a dark mode with respect to the fiber, provided the QST is performed at short enough ranges, see the discussion at the end of Sec.~\ref{ssec:trieff}.

To fully investigate the effect of this leakage of the information we perform a numerical analysis on the adiabatic passage protocol.
The numerics of the SAP protocol are too expensive for the multimode fiber case.
Instead, we shall elaborate an effective single-mode model that includes the dynamics of the higher order modes through their adiabatic elimination in the next section.

\begin{figure}[tb]
\includegraphics[width=\columnwidth]{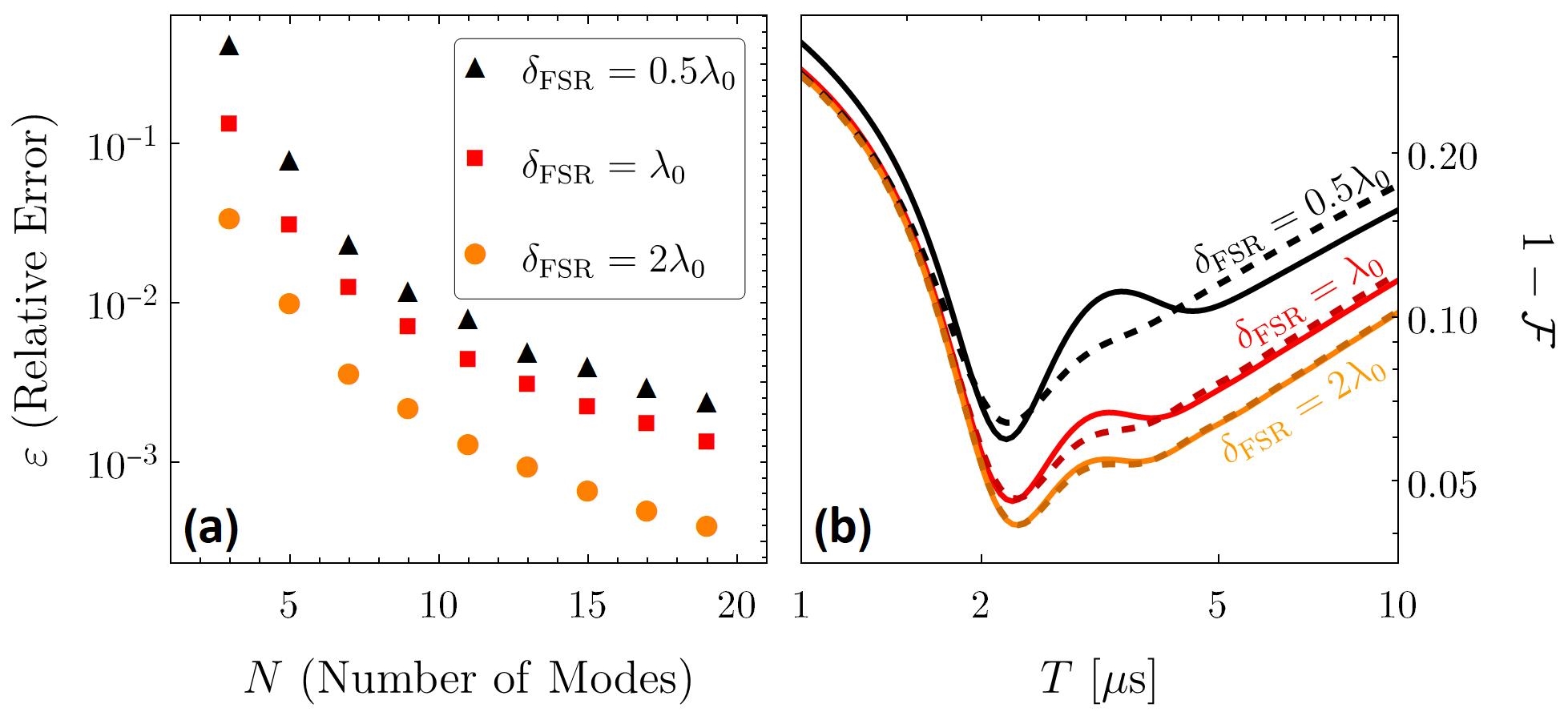}
\caption{%
(a) Variations of the relative error in the final fidelity $\varepsilon$ with respect to the number of fiber modes included in the numerics for three different values of $\fsr$. Here, $T=2000$~ns and the transferred state is $\ket{\alpha=1,r=1}$.
(b) The infidelity in QST of the same state as a function of the pulse duration for three different free spectral range values. The number of modes included in the calculations are chosen such that the relative error is less than one percent.
The dashed lines show results from the effective single-mode model.
The other parameters used in both panels are $\lambda_0/2\pi=10$~MHz and $n_{\rm th}=0$.
}%
\label{fig:multiap}%
\end{figure}

The numerical results are presented in Fig.~\ref{fig:multiap} for the QST of $\ket{\alpha=1,r=1}$.
Obviously, due to the limits in the computational resources only a finite number of fiber modes can be taken into account.
Hence, we first analyze the error that one commits by this consideration. That is, fidelity of the adiabatic passage protocol is computed for different number of fiber modes $N=1,3,5,\cdots$, where the modes nearest to the resonance mode $f_0$ are symmetrically included at each stage.
The relative error is then calculated by
\begin{equation}
 \varepsilon_N \equiv 2\frac{|\mathcal{F}_{N} -\mathcal{F}_{N-2}|}{\mathcal{F}_{N} +\mathcal{F}_{N-2}},\hspace{10mm}(N \geq 3)
\end{equation}
where $\mathcal{F}_N$ is the fidelity of an $N$-mode fiber with mode indices $n\in[-\half(N-1),+\half(N-1)]$ involved in the computations.
We plot the relative errors in Fig.~\ref{fig:multiap}(a) for three different free spectral range values.
A monotonic decrease in the error suggests that the fidelity converges rather rapidly as the number of involved fiber modes increases.
The other important observation is that as $\fsr$ grows greater than $\lambda_0$, contribution of the farther fiber modes in the QST process becomes negligible.
For example, when $\fsr=2\lambda_0$ the relative error is already less than five percent by only adding $f_{\pm 1}$ modes into the calculations.

We next investigate performance of the protocol with respect to the free spectral range, i.e., length of the fiber. The representative results are shown in Fig.~\ref{fig:multiap}(b), where each curve corresponds to the infidelity of the protocol for a given value of $\fsr$ as a function of the protocol duration $T$.
In performing the numerics we have included enough number of fiber modes into the calculations such that a relative error less than one percent is guaranteed. That is, for $\fsr=0.5\lambda_0$ thirteen fiber modes with indices $n\in[-6,+6]$, $\fsr=\lambda_0$ nine fiber modes with $n\in[-4,+4]$, and for $\fsr=2\lambda_0$ only five fiber modes with $n\in[-2,+2]$ are involved in the calculations.
The clear message of the plot is that the information leakage to the higher fiber modes is suppressed, provided the fiber free spectral range dominates the rate of coupling of the mechanical nodes to the fiber.
This result is in agreement with Eq.~\eqref{darkmode} and the discussion below it. That is, for $\fsr > g_1,g_2$ the chance of exciting higher fiber modes becomes negligible, and thus, the dark adiabatic mode is the main state transfer channel.
It is worth reminding that to ensure the validity of cavity adiabatic elimination, we have adjusted the branching ratio such that $\chi=0.2\kappa_0$.
Therefore, we compute $\eta \approx \{0.557,0.386,0.239\}$ and $\kappa \approx\{22.6,8.1,3.3\}\fsr$ for $\fsr=\{0.5,1,2\}\lambda_0$, respectively.

%
\subsection{Effective single-mode model\label{ssec:trieff}}
The numerical analysis of the multimode fiber case is too expensive for the shortcut to adiabatic passage protocol. Therefore, we elaborate an effective single-mode fiber model by adiabatic elimination of the higher order modes and only keeping the nearest to the resonance mode $f_0$.

\begin{figure}[tb]
\includegraphics[width=\columnwidth]{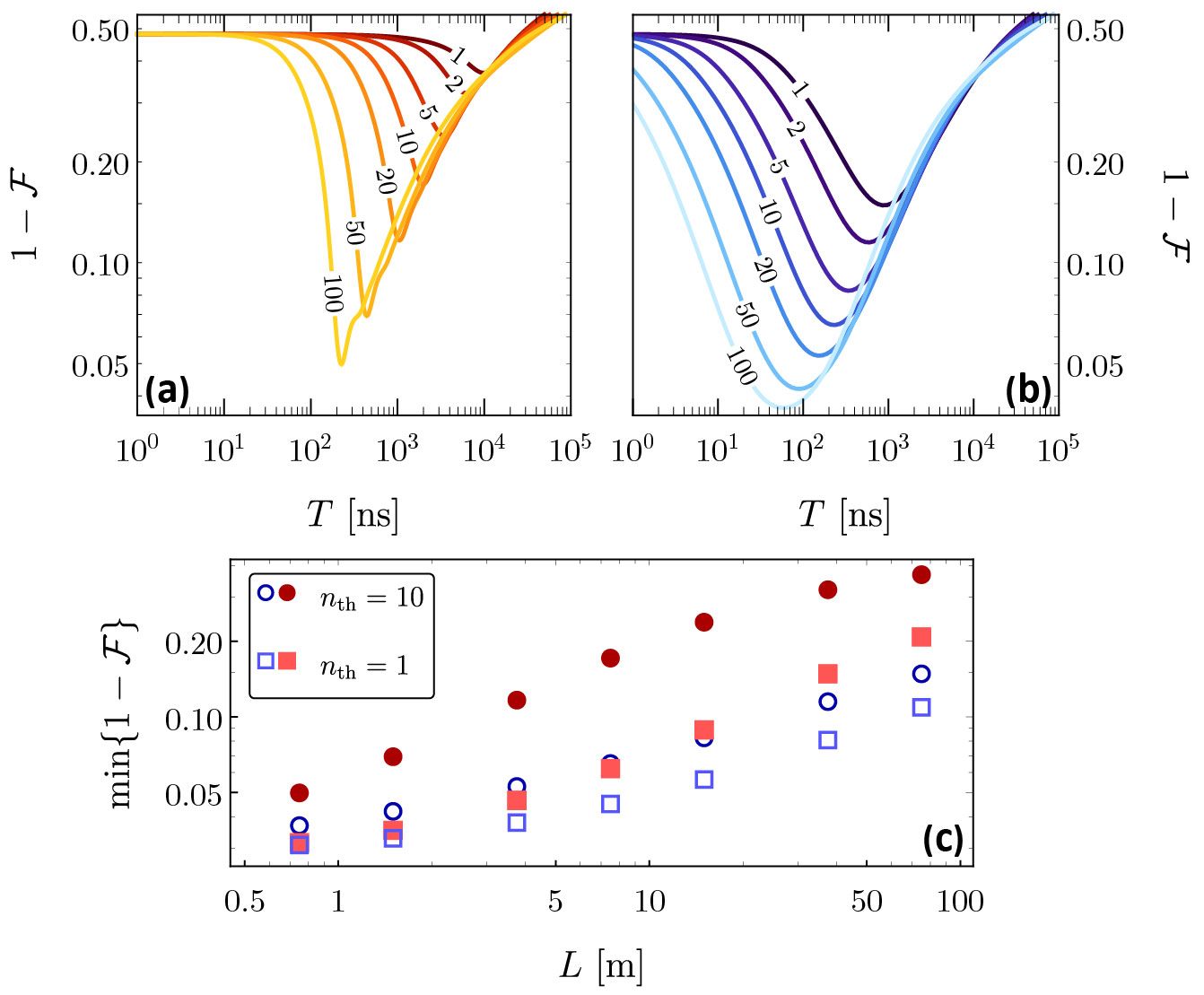}
\caption{%
Effective single-mode model:
Infidelity of the AP (a) and SAP (b) QST protocols as a function of the protocol duration $T$ for various coupling rate values. The number on the curves give $\lambda_0/2\pi$ in units of megahertz. The thermal occupation number is set to $n_{\rm th}=10$.
(c) The optimal performance (minimum infidelity) of the AP (filled markers) and SAP (open markers) with respect to the length of the fiber $L=\pi c/\fsr$ for two different mechanical bath thermal occupation numbers.
The state under study is the coherent squeezed state $\ket{1,1}$, $\fsr=2\lambda_0$.
Here, $\eta \approx 0.24$ and similar to the rest of paper we have set $\kappa_0=5\lambda_0$.
}%
\label{fig:multistap}%
\end{figure}

The dynamics of $n$th fiber mode can be adiabatically eliminated if the condition $\widetilde\gamma_f^{2} +n^2\fsr^{2} \gtrsim \widetilde\gamma_f^2,\widetilde\gamma_{1(2)},g_{i,n}^{2}$ is satisfied.
Considering that we are interested in the regime where $\kappa_0 \gg \lambda_0,\chi$ this condition simplifies to $n\fsr \gtrsim \lambda_0$ for all fiber modes with $n\neq 0$.
Therefore, in the following we eliminate all fiber modes except for the `resonant' mode by assuming $\fsr \gtrsim \lambda_0$.
For this, we set $\dot{f}_{n\neq 0}=0$ in the left hand side of Eq.~\eqref{efflangevinb}, and then find the steady state relation of $f_{n\neq 0}$ operators.
By plugging them back into the equations for $m_1$, $m_2$, and $f_0$ an effective three-mode system dynamics is derived.
Hence, one arrives at a set of equations as in Eq.~\eqref{heisenberg} but the dynamics matrix \eqref{dynamics} replaced by
\begin{equation}
M(t) = \left(
\begin{array}{ccc}
-i(\widetilde\gamma_1+\Gamma_{11}) & g_1 & -i\Gamma_{12} \\
g_1 & -\widetilde\gamma_f & g_2 \\
-i\Gamma_{12} & g_2 & -i(\widetilde\gamma_2 +\Gamma_{22})
\end{array}\right),
\label{dynamiceff}
\end{equation}
where $\Gamma_{ij} \equiv \widetilde\gamma_f \sum_{n\neq 0}\frac{g_{i,n}g_{j,n}}{\widetilde\gamma_f^2 +n^2\fsr^2}$ 
with $(i,j=1,2)$ are the fiber-induced mechanical damping rates.
The cross damping $\Gamma_{12}$ induces a dissipative coupling between the mechanical nodes.
Meanwhile, the noise operator vector is modified to $n = (\Xi_1, \varphi_0, \Xi_2)^\intercal$ with the modified mechanical noise operators given by
\begin{equation}
\Xi_i \equiv \xi_i -i\sum_{n\neq 0}\frac{g_{i,n}}{\widetilde\gamma_f +in\fsr}\varphi_n.
\label{noiseeff}%
\end{equation}%
Alongside the modification of the diagonal elements in the effective diffusion matrix, these effective noise operators give rise to cross-mechanical mode noise correlations.
The results of this effective model in the adiabatic passage protocol are compared to the exact multimode results in Fig.~\ref{fig:multiap}(b).
One infers a reasonable agreement between the two results provided the free spectral range satisfies the condition $\fsr \gtrsim \lambda_0$.
This is such that already for $\fsr=2\lambda_0$ the effective model captures almost all features of the exact numerical results.
Having established the validity of the effective single-mode model we now move to study performance of the SAP protocol in a multimode fiber.

In our numerical analysis with the effective single-mode fiber model we set $\fsr=2\lambda_0$ to ensure the validity of the results.
The fidelity of the QST via AP and SAP protocols are investigated for different coupling rates $\lambda_0$, which as for the fixed free spectral range value correspond to different multimode fiber lengths $L$.
The results are presented in Fig.~\ref{fig:multistap}, where the infidelities are plotted against duration of the protocols for various values of coupling rate $\lambda_0$.
As the coupling rate increases, the performance of both protocols is increased and the required pulse duration becomes smaller, but the SAP is always faster and more efficient [Fig.~\ref{fig:multistap}(a) and (b)].
The overall performance of both protocols is almost the same at low temperatures and short fiber lengths.
Nevertheless, for high mechanical bath occupation numbers the SAP outperforms the AP protocol.
At any temperature, the performance drops by increasing the length of fiber, evidently because of the reduced coupling rate $\lambda_0$, see Fig.~\ref{fig:multistap}(c).
This in turn stems from the necessity of satisfying the condition $\fsr \gtrsim \lambda_0$, which guarantees transfer of the quantum state through dark fiber mode when a multimode fiber is employed.

%
%
\section{Conclusion}\label{sec:conclusion}
We have investigated the quantum state transfer between two distant mechanical resonators coupled via a lossy optical fiber. Specifically, two protocols are studied: the adiabatic passage and shortcut to adiabatic passage.
By the shortcut to adiabatic passage protocol one eliminates the diabatic transitions between adiabatic modes of the system during the transfer process.
We have shown that this can be done by the modification of the coupling pulses and that it results-in a fast and efficient state transfer.
The effect of thermal noise in the local nodes on the efficiency of the protocols has been investigated, showing that the fast functioning of the SAP protocol is crucial in retaining the QST performance in high temperatures.
The case of multimode fiber has also been considered via both numerical calculations and an effective single-mode model that relies on adiabatic elimination of the off-resonant fiber modes.
We find that a safe QST requires a large free spectral range for the fiber.

\begin{acknowledgments}
MA acknowledges the support by Iran Science Elites Federation.
\end{acknowledgments}

\bibliography{QST_3.1.bbl}

\begin{thebibliography}{66}%
\makeatletter
\providecommand \@ifxundefined [1]{%
 \@ifx{#1\undefined}
}%
\providecommand \@ifnum [1]{%
 \ifnum #1\expandafter \@firstoftwo
 \else \expandafter \@secondoftwo
 \fi
}%
\providecommand \@ifx [1]{%
 \ifx #1\expandafter \@firstoftwo
 \else \expandafter \@secondoftwo
 \fi
}%
\providecommand \natexlab [1]{#1}%
\providecommand \enquote  [1]{``#1''}%
\providecommand \bibnamefont  [1]{#1}%
\providecommand \bibfnamefont [1]{#1}%
\providecommand \citenamefont [1]{#1}%
\providecommand \href@noop [0]{\@secondoftwo}%
\providecommand \href [0]{\begingroup \@sanitize@url \@href}%
\providecommand \@href[1]{\@@startlink{#1}\@@href}%
\providecommand \@@href[1]{\endgroup#1\@@endlink}%
\providecommand \@sanitize@url [0]{\catcode `\\12\catcode `\$12\catcode
  `\&12\catcode `\#12\catcode `\^12\catcode `\_12\catcode `\%12\relax}%
\providecommand \@@startlink[1]{}%
\providecommand \@@endlink[0]{}%
\providecommand \url  [0]{\begingroup\@sanitize@url \@url }%
\providecommand \@url [1]{\endgroup\@href {#1}{\urlprefix }}%
\providecommand \urlprefix  [0]{URL }%
\providecommand \Eprint [0]{\href }%
\providecommand \doibase [0]{https://doi.org/}%
\providecommand \selectlanguage [0]{\@gobble}%
\providecommand \bibinfo  [0]{\@secondoftwo}%
\providecommand \bibfield  [0]{\@secondoftwo}%
\providecommand \translation [1]{[#1]}%
\providecommand \BibitemOpen [0]{}%
\providecommand \bibitemStop [0]{}%
\providecommand \bibitemNoStop [0]{.\EOS\space}%
\providecommand \EOS [0]{\spacefactor3000\relax}%
\providecommand \BibitemShut  [1]{\csname bibitem#1\endcsname}%
\let\auto@bib@innerbib\@empty
\bibitem [{\citenamefont {Kimble}(2008)}]{kimble2008quantum}%
  \BibitemOpen
  \bibfield  {author} {\bibinfo {author} {\bibfnamefont {H.~J.}\ \bibnamefont
  {Kimble}},\ }\bibfield  {title} {\bibinfo {title} {The quantum internet},\
  }\href {https://doi.org/10.1038/nature07127} {\bibfield  {journal} {\bibinfo
  {journal} {Nature}\ }\textbf {\bibinfo {volume} {453}},\ \bibinfo {pages}
  {1023} (\bibinfo {year} {2008})}\BibitemShut {NoStop}%
\bibitem [{\citenamefont {Northup}\ and\ \citenamefont
  {Blatt}(2014)}]{northup2014quantum}%
  \BibitemOpen
  \bibfield  {author} {\bibinfo {author} {\bibfnamefont {T.}~\bibnamefont
  {Northup}}\ and\ \bibinfo {author} {\bibfnamefont {R.}~\bibnamefont
  {Blatt}},\ }\bibfield  {title} {\bibinfo {title} {Quantum information
  transfer using photons},\ }\href {https://doi.org/10.1038/nphoton.2014.53}
  {\bibfield  {journal} {\bibinfo  {journal} {Nat. Photonics}\ }\textbf
  {\bibinfo {volume} {8}},\ \bibinfo {pages} {356} (\bibinfo {year}
  {2014})}\BibitemShut {NoStop}%
\bibitem [{\citenamefont {Hammerer}\ \emph {et~al.}(2010)\citenamefont
  {Hammerer}, \citenamefont {S{\o}rensen},\ and\ \citenamefont
  {Polzik}}]{hammerer2010quantum}%
  \BibitemOpen
  \bibfield  {author} {\bibinfo {author} {\bibfnamefont {K.}~\bibnamefont
  {Hammerer}}, \bibinfo {author} {\bibfnamefont {A.~S.}\ \bibnamefont
  {S{\o}rensen}},\ and\ \bibinfo {author} {\bibfnamefont {E.~S.}\ \bibnamefont
  {Polzik}},\ }\bibfield  {title} {\bibinfo {title} {Quantum interface between
  light and atomic ensembles},\ }\href
  {https://doi.org/10.1103/RevModPhys.82.1041} {\bibfield  {journal} {\bibinfo
  {journal} {Rev. Mod. Phys.}\ }\textbf {\bibinfo {volume} {82}},\ \bibinfo
  {pages} {1041} (\bibinfo {year} {2010})}\BibitemShut {NoStop}%
\bibitem [{\citenamefont {Bennett}\ \emph {et~al.}(1993)\citenamefont
  {Bennett}, \citenamefont {Brassard}, \citenamefont {Cr{\'{e}}peau},
  \citenamefont {Jozsa}, \citenamefont {Peres},\ and\ \citenamefont
  {Wootters}}]{Bennett1993}%
  \BibitemOpen
  \bibfield  {author} {\bibinfo {author} {\bibfnamefont {C.~H.}\ \bibnamefont
  {Bennett}}, \bibinfo {author} {\bibfnamefont {G.}~\bibnamefont {Brassard}},
  \bibinfo {author} {\bibfnamefont {C.}~\bibnamefont {Cr{\'{e}}peau}}, \bibinfo
  {author} {\bibfnamefont {R.}~\bibnamefont {Jozsa}}, \bibinfo {author}
  {\bibfnamefont {A.}~\bibnamefont {Peres}},\ and\ \bibinfo {author}
  {\bibfnamefont {W.~K.}\ \bibnamefont {Wootters}},\ }\bibfield  {title}
  {\bibinfo {title} {Teleporting an unknown quantum state via dual classical
  and einstein-podolsky-rosen channels},\ }\href
  {https://doi.org/10.1103/physrevlett.70.1895} {\bibfield  {journal} {\bibinfo
   {journal} {Phys. Rev. Lett.}\ }\textbf {\bibinfo {volume} {70}},\ \bibinfo
  {pages} {1895} (\bibinfo {year} {1993})}\BibitemShut {NoStop}%
\bibitem [{\citenamefont {Braunstein}\ and\ \citenamefont
  {Kimble}(1998)}]{Braunstein1998}%
  \BibitemOpen
  \bibfield  {author} {\bibinfo {author} {\bibfnamefont {S.~L.}\ \bibnamefont
  {Braunstein}}\ and\ \bibinfo {author} {\bibfnamefont {H.~J.}\ \bibnamefont
  {Kimble}},\ }\bibfield  {title} {\bibinfo {title} {Teleportation of
  continuous quantum variables},\ }\href
  {https://doi.org/10.1103/physrevlett.80.869} {\bibfield  {journal} {\bibinfo
  {journal} {Phys. Rev. Lett.}\ }\textbf {\bibinfo {volume} {80}},\ \bibinfo
  {pages} {869} (\bibinfo {year} {1998})}\BibitemShut {NoStop}%
\bibitem [{\citenamefont {Bienfait}\ \emph {et~al.}(2019)\citenamefont
  {Bienfait}, \citenamefont {Satzinger}, \citenamefont {Zhong}, \citenamefont
  {Chang}, \citenamefont {Chou}, \citenamefont {Conner}, \citenamefont {Dumur},
  \citenamefont {Grebel}, \citenamefont {Peairs}, \citenamefont {Povey},\ and\
  \citenamefont {Cleland}}]{Bienfait2019}%
  \BibitemOpen
  \bibfield  {author} {\bibinfo {author} {\bibfnamefont {A.}~\bibnamefont
  {Bienfait}}, \bibinfo {author} {\bibfnamefont {K.~J.}\ \bibnamefont
  {Satzinger}}, \bibinfo {author} {\bibfnamefont {Y.~P.}\ \bibnamefont
  {Zhong}}, \bibinfo {author} {\bibfnamefont {H.-S.}\ \bibnamefont {Chang}},
  \bibinfo {author} {\bibfnamefont {M.-H.}\ \bibnamefont {Chou}}, \bibinfo
  {author} {\bibfnamefont {C.~R.}\ \bibnamefont {Conner}}, \bibinfo {author}
  {\bibfnamefont {{\'{E}}.}~\bibnamefont {Dumur}}, \bibinfo {author}
  {\bibfnamefont {J.}~\bibnamefont {Grebel}}, \bibinfo {author} {\bibfnamefont
  {G.~A.}\ \bibnamefont {Peairs}}, \bibinfo {author} {\bibfnamefont {R.~G.}\
  \bibnamefont {Povey}},\ and\ \bibinfo {author} {\bibfnamefont {A.~N.}\
  \bibnamefont {Cleland}},\ }\bibfield  {title} {\bibinfo {title}
  {Phonon-mediated quantum state transfer and remote qubit entanglement},\
  }\href {https://doi.org/10.1126/science.aaw8415} {\bibfield  {journal}
  {\bibinfo  {journal} {Science}\ }\textbf {\bibinfo {volume} {364}},\ \bibinfo
  {pages} {368} (\bibinfo {year} {2019})}\BibitemShut {NoStop}%
\bibitem [{\citenamefont {Kurpiers}\ \emph {et~al.}(2018)\citenamefont
  {Kurpiers}, \citenamefont {Magnard}, \citenamefont {Walter}, \citenamefont
  {Royer}, \citenamefont {Pechal}, \citenamefont {Heinsoo}, \citenamefont
  {Salath{\'{e}}}, \citenamefont {Akin}, \citenamefont {Storz}, \citenamefont
  {Besse}, \citenamefont {Gasparinetti}, \citenamefont {Blais},\ and\
  \citenamefont {Wallraff}}]{Kurpiers2018}%
  \BibitemOpen
  \bibfield  {author} {\bibinfo {author} {\bibfnamefont {P.}~\bibnamefont
  {Kurpiers}}, \bibinfo {author} {\bibfnamefont {P.}~\bibnamefont {Magnard}},
  \bibinfo {author} {\bibfnamefont {T.}~\bibnamefont {Walter}}, \bibinfo
  {author} {\bibfnamefont {B.}~\bibnamefont {Royer}}, \bibinfo {author}
  {\bibfnamefont {M.}~\bibnamefont {Pechal}}, \bibinfo {author} {\bibfnamefont
  {J.}~\bibnamefont {Heinsoo}}, \bibinfo {author} {\bibfnamefont
  {Y.}~\bibnamefont {Salath{\'{e}}}}, \bibinfo {author} {\bibfnamefont
  {A.}~\bibnamefont {Akin}}, \bibinfo {author} {\bibfnamefont {S.}~\bibnamefont
  {Storz}}, \bibinfo {author} {\bibfnamefont {J.-C.}\ \bibnamefont {Besse}},
  \bibinfo {author} {\bibfnamefont {S.}~\bibnamefont {Gasparinetti}}, \bibinfo
  {author} {\bibfnamefont {A.}~\bibnamefont {Blais}},\ and\ \bibinfo {author}
  {\bibfnamefont {A.}~\bibnamefont {Wallraff}},\ }\bibfield  {title} {\bibinfo
  {title} {Deterministic quantum state transfer and remote entanglement using
  microwave photons},\ }\href {https://doi.org/10.1038/s41586-018-0195-y}
  {\bibfield  {journal} {\bibinfo  {journal} {Nature}\ }\textbf {\bibinfo
  {volume} {558}},\ \bibinfo {pages} {264} (\bibinfo {year}
  {2018})}\BibitemShut {NoStop}%
\bibitem [{\citenamefont {Vermersch}\ \emph {et~al.}(2017)\citenamefont
  {Vermersch}, \citenamefont {Guimond}, \citenamefont {Pichler},\ and\
  \citenamefont {Zoller}}]{Vermersch2017}%
  \BibitemOpen
  \bibfield  {author} {\bibinfo {author} {\bibfnamefont {B.}~\bibnamefont
  {Vermersch}}, \bibinfo {author} {\bibfnamefont {P.-O.}\ \bibnamefont
  {Guimond}}, \bibinfo {author} {\bibfnamefont {H.}~\bibnamefont {Pichler}},\
  and\ \bibinfo {author} {\bibfnamefont {P.}~\bibnamefont {Zoller}},\
  }\bibfield  {title} {\bibinfo {title} {Quantum state transfer via noisy
  photonic and phononic waveguides},\ }\href
  {https://doi.org/10.1103/physrevlett.118.133601} {\bibfield  {journal}
  {\bibinfo  {journal} {Phys. Rev. Lett.}\ }\textbf {\bibinfo {volume} {118}},\
  \bibinfo {pages} {133601} (\bibinfo {year} {2017})}\BibitemShut {NoStop}%
\bibitem [{\citenamefont {Ursin}\ \emph {et~al.}(2007)\citenamefont {Ursin},
  \citenamefont {Tiefenbacher}, \citenamefont {Schmitt-Manderbach},
  \citenamefont {Weier}, \citenamefont {Scheidl}, \citenamefont {Lindenthal},
  \citenamefont {Blauensteiner}, \citenamefont {Jennewein}, \citenamefont
  {Perdigues}, \citenamefont {Trojek} \emph {et~al.}}]{ursin2007entanglement}%
  \BibitemOpen
  \bibfield  {author} {\bibinfo {author} {\bibfnamefont {R.}~\bibnamefont
  {Ursin}}, \bibinfo {author} {\bibfnamefont {F.}~\bibnamefont {Tiefenbacher}},
  \bibinfo {author} {\bibfnamefont {T.}~\bibnamefont {Schmitt-Manderbach}},
  \bibinfo {author} {\bibfnamefont {H.}~\bibnamefont {Weier}}, \bibinfo
  {author} {\bibfnamefont {T.}~\bibnamefont {Scheidl}}, \bibinfo {author}
  {\bibfnamefont {M.}~\bibnamefont {Lindenthal}}, \bibinfo {author}
  {\bibfnamefont {B.}~\bibnamefont {Blauensteiner}}, \bibinfo {author}
  {\bibfnamefont {T.}~\bibnamefont {Jennewein}}, \bibinfo {author}
  {\bibfnamefont {J.}~\bibnamefont {Perdigues}}, \bibinfo {author}
  {\bibfnamefont {P.}~\bibnamefont {Trojek}}, \emph {et~al.},\ }\bibfield
  {title} {\bibinfo {title} {Entanglement-based quantum communication over 144
  km},\ }\href {https://doi.org/10.1038/nphys629} {\bibfield  {journal}
  {\bibinfo  {journal} {Nat. Phys.}\ }\textbf {\bibinfo {volume} {3}},\
  \bibinfo {pages} {481} (\bibinfo {year} {2007})}\BibitemShut {NoStop}%
\bibitem [{\citenamefont {Ritter}\ \emph {et~al.}(2012)\citenamefont {Ritter},
  \citenamefont {N{\"o}lleke}, \citenamefont {Hahn}, \citenamefont {Reiserer},
  \citenamefont {Neuzner}, \citenamefont {Uphoff}, \citenamefont {M{\"u}cke},
  \citenamefont {Figueroa}, \citenamefont {Bochmann},\ and\ \citenamefont
  {Rempe}}]{ritter2012elementary}%
  \BibitemOpen
  \bibfield  {author} {\bibinfo {author} {\bibfnamefont {S.}~\bibnamefont
  {Ritter}}, \bibinfo {author} {\bibfnamefont {C.}~\bibnamefont {N{\"o}lleke}},
  \bibinfo {author} {\bibfnamefont {C.}~\bibnamefont {Hahn}}, \bibinfo {author}
  {\bibfnamefont {A.}~\bibnamefont {Reiserer}}, \bibinfo {author}
  {\bibfnamefont {A.}~\bibnamefont {Neuzner}}, \bibinfo {author} {\bibfnamefont
  {M.}~\bibnamefont {Uphoff}}, \bibinfo {author} {\bibfnamefont
  {M.}~\bibnamefont {M{\"u}cke}}, \bibinfo {author} {\bibfnamefont
  {E.}~\bibnamefont {Figueroa}}, \bibinfo {author} {\bibfnamefont
  {J.}~\bibnamefont {Bochmann}},\ and\ \bibinfo {author} {\bibfnamefont
  {G.}~\bibnamefont {Rempe}},\ }\bibfield  {title} {\bibinfo {title} {An
  elementary quantum network of single atoms in optical cavities},\ }\href
  {https://doi.org/10.1038/nature11023} {\bibfield  {journal} {\bibinfo
  {journal} {Nature}\ }\textbf {\bibinfo {volume} {484}},\ \bibinfo {pages}
  {195} (\bibinfo {year} {2012})}\BibitemShut {NoStop}%
\bibitem [{\citenamefont {Olmschenk}\ \emph {et~al.}(2009)\citenamefont
  {Olmschenk}, \citenamefont {Matsukevich}, \citenamefont {Maunz},
  \citenamefont {Hayes}, \citenamefont {Duan},\ and\ \citenamefont
  {Monroe}}]{Olmschenk2009}%
  \BibitemOpen
  \bibfield  {author} {\bibinfo {author} {\bibfnamefont {S.}~\bibnamefont
  {Olmschenk}}, \bibinfo {author} {\bibfnamefont {D.~N.}\ \bibnamefont
  {Matsukevich}}, \bibinfo {author} {\bibfnamefont {P.}~\bibnamefont {Maunz}},
  \bibinfo {author} {\bibfnamefont {D.}~\bibnamefont {Hayes}}, \bibinfo
  {author} {\bibfnamefont {L.-M.}\ \bibnamefont {Duan}},\ and\ \bibinfo
  {author} {\bibfnamefont {C.}~\bibnamefont {Monroe}},\ }\bibfield  {title}
  {\bibinfo {title} {Quantum teleportation between distant matter qubits},\
  }\href {https://doi.org/10.1126/science.1167209} {\bibfield  {journal}
  {\bibinfo  {journal} {Science}\ }\textbf {\bibinfo {volume} {323}},\ \bibinfo
  {pages} {486} (\bibinfo {year} {2009})}\BibitemShut {NoStop}%
\bibitem [{\citenamefont {Axline}\ \emph {et~al.}(2018)\citenamefont {Axline},
  \citenamefont {Burkhart}, \citenamefont {Pfaff}, \citenamefont {Zhang},
  \citenamefont {Chou}, \citenamefont {Campagne-Ibarcq}, \citenamefont
  {Reinhold}, \citenamefont {Frunzio}, \citenamefont {Girvin}, \citenamefont
  {Jiang}, \citenamefont {Devoret},\ and\ \citenamefont
  {Schoelkopf}}]{Axline2018}%
  \BibitemOpen
  \bibfield  {author} {\bibinfo {author} {\bibfnamefont {C.~J.}\ \bibnamefont
  {Axline}}, \bibinfo {author} {\bibfnamefont {L.~D.}\ \bibnamefont
  {Burkhart}}, \bibinfo {author} {\bibfnamefont {W.}~\bibnamefont {Pfaff}},
  \bibinfo {author} {\bibfnamefont {M.}~\bibnamefont {Zhang}}, \bibinfo
  {author} {\bibfnamefont {K.}~\bibnamefont {Chou}}, \bibinfo {author}
  {\bibfnamefont {P.}~\bibnamefont {Campagne-Ibarcq}}, \bibinfo {author}
  {\bibfnamefont {P.}~\bibnamefont {Reinhold}}, \bibinfo {author}
  {\bibfnamefont {L.}~\bibnamefont {Frunzio}}, \bibinfo {author} {\bibfnamefont
  {S.~M.}\ \bibnamefont {Girvin}}, \bibinfo {author} {\bibfnamefont
  {L.}~\bibnamefont {Jiang}}, \bibinfo {author} {\bibfnamefont {M.~H.}\
  \bibnamefont {Devoret}},\ and\ \bibinfo {author} {\bibfnamefont {R.~J.}\
  \bibnamefont {Schoelkopf}},\ }\bibfield  {title} {\bibinfo {title} {On-demand
  quantum state transfer and entanglement between remote microwave cavity
  memories},\ }\href {https://doi.org/10.1038/s41567-018-0115-y} {\bibfield
  {journal} {\bibinfo  {journal} {Nat. Phys.}\ }\textbf {\bibinfo {volume}
  {14}},\ \bibinfo {pages} {705} (\bibinfo {year} {2018})}\BibitemShut
  {NoStop}%
\bibitem [{\citenamefont {Cirac}\ \emph {et~al.}(1997)\citenamefont {Cirac},
  \citenamefont {Zoller}, \citenamefont {Kimble},\ and\ \citenamefont
  {Mabuchi}}]{cirac1997quantum}%
  \BibitemOpen
  \bibfield  {author} {\bibinfo {author} {\bibfnamefont {J.~I.}\ \bibnamefont
  {Cirac}}, \bibinfo {author} {\bibfnamefont {P.}~\bibnamefont {Zoller}},
  \bibinfo {author} {\bibfnamefont {H.~J.}\ \bibnamefont {Kimble}},\ and\
  \bibinfo {author} {\bibfnamefont {H.}~\bibnamefont {Mabuchi}},\ }\bibfield
  {title} {\bibinfo {title} {Quantum state transfer and entanglement
  distribution among distant nodes in a quantum network},\ }\href
  {https://doi.org/10.1103/PhysRevLett.78.3221} {\bibfield  {journal} {\bibinfo
   {journal} {Phys. Rev. Lett.}\ }\textbf {\bibinfo {volume} {78}},\ \bibinfo
  {pages} {3221} (\bibinfo {year} {1997})}\BibitemShut {NoStop}%
\bibitem [{\citenamefont {Dantan}\ and\ \citenamefont
  {Pinard}(2004)}]{Dantan2004}%
  \BibitemOpen
  \bibfield  {author} {\bibinfo {author} {\bibfnamefont {A.}~\bibnamefont
  {Dantan}}\ and\ \bibinfo {author} {\bibfnamefont {M.}~\bibnamefont
  {Pinard}},\ }\bibfield  {title} {\bibinfo {title} {Quantum-state transfer
  between fields and atoms in electromagnetically induced transparency},\
  }\href {https://doi.org/10.1103/physreva.69.043810} {\bibfield  {journal}
  {\bibinfo  {journal} {Phys. Rev. A}\ }\textbf {\bibinfo {volume} {69}},\
  \bibinfo {pages} {043810} (\bibinfo {year} {2004})}\BibitemShut {NoStop}%
\bibitem [{\citenamefont {Vogell}\ \emph {et~al.}(2017)\citenamefont {Vogell},
  \citenamefont {Vermersch}, \citenamefont {Northup}, \citenamefont {Lanyon},\
  and\ \citenamefont {Muschik}}]{Vogell2017}%
  \BibitemOpen
  \bibfield  {author} {\bibinfo {author} {\bibfnamefont {B.}~\bibnamefont
  {Vogell}}, \bibinfo {author} {\bibfnamefont {B.}~\bibnamefont {Vermersch}},
  \bibinfo {author} {\bibfnamefont {T.~E.}\ \bibnamefont {Northup}}, \bibinfo
  {author} {\bibfnamefont {B.~P.}\ \bibnamefont {Lanyon}},\ and\ \bibinfo
  {author} {\bibfnamefont {C.~A.}\ \bibnamefont {Muschik}},\ }\bibfield
  {title} {\bibinfo {title} {Deterministic quantum state transfer between
  remote qubits in cavities},\ }\href
  {https://doi.org/10.1088/2058-9565/aa868b} {\bibfield  {journal} {\bibinfo
  {journal} {Quantum Sci. Technol.}\ }\textbf {\bibinfo {volume} {2}},\
  \bibinfo {pages} {045003} (\bibinfo {year} {2017})}\BibitemShut {NoStop}%
\bibitem [{\citenamefont {Vitanov}\ and\ \citenamefont
  {Stenholm}(1997)}]{vitanov1997analytic}%
  \BibitemOpen
  \bibfield  {author} {\bibinfo {author} {\bibfnamefont {N.~V.}\ \bibnamefont
  {Vitanov}}\ and\ \bibinfo {author} {\bibfnamefont {S.}~\bibnamefont
  {Stenholm}},\ }\bibfield  {title} {\bibinfo {title} {Analytic properties and
  effective two-level problems in stimulated raman adiabatic passage},\ }\href
  {https://doi.org/10.1103/PhysRevA.55.648} {\bibfield  {journal} {\bibinfo
  {journal} {Phys. Rev. A}\ }\textbf {\bibinfo {volume} {55}},\ \bibinfo
  {pages} {648} (\bibinfo {year} {1997})}\BibitemShut {NoStop}%
\bibitem [{\citenamefont {Braunstein}\ and\ \citenamefont {van
  Loock}(2005)}]{Braunstein2005}%
  \BibitemOpen
  \bibfield  {author} {\bibinfo {author} {\bibfnamefont {S.~L.}\ \bibnamefont
  {Braunstein}}\ and\ \bibinfo {author} {\bibfnamefont {P.}~\bibnamefont {van
  Loock}},\ }\bibfield  {title} {\bibinfo {title} {Quantum information with
  continuous variables},\ }\href {https://doi.org/10.1103/revmodphys.77.513}
  {\bibfield  {journal} {\bibinfo  {journal} {Rev. Mod. Phys.}\ }\textbf
  {\bibinfo {volume} {77}},\ \bibinfo {pages} {513} (\bibinfo {year}
  {2005})}\BibitemShut {NoStop}%
\bibitem [{\citenamefont {Stannigel}\ \emph {et~al.}(2010)\citenamefont
  {Stannigel}, \citenamefont {Rabl}, \citenamefont {S{\o}rensen}, \citenamefont
  {Zoller},\ and\ \citenamefont {Lukin}}]{stannigel2010optomechanical}%
  \BibitemOpen
  \bibfield  {author} {\bibinfo {author} {\bibfnamefont {K.}~\bibnamefont
  {Stannigel}}, \bibinfo {author} {\bibfnamefont {P.}~\bibnamefont {Rabl}},
  \bibinfo {author} {\bibfnamefont {A.~S.}\ \bibnamefont {S{\o}rensen}},
  \bibinfo {author} {\bibfnamefont {P.}~\bibnamefont {Zoller}},\ and\ \bibinfo
  {author} {\bibfnamefont {M.~D.}\ \bibnamefont {Lukin}},\ }\bibfield  {title}
  {\bibinfo {title} {Optomechanical transducers for long-distance quantum
  communication},\ }\href {https://doi.org/10.1103/PhysRevLett.105.220501}
  {\bibfield  {journal} {\bibinfo  {journal} {Phys. Rev. Lett.}\ }\textbf
  {\bibinfo {volume} {105}},\ \bibinfo {pages} {220501} (\bibinfo {year}
  {2010})}\BibitemShut {NoStop}%
\bibitem [{\citenamefont {McGee}\ \emph {et~al.}(2013)\citenamefont {McGee},
  \citenamefont {Meiser}, \citenamefont {Regal}, \citenamefont {Lehnert},\ and\
  \citenamefont {Holland}}]{McGee2013}%
  \BibitemOpen
  \bibfield  {author} {\bibinfo {author} {\bibfnamefont {S.~A.}\ \bibnamefont
  {McGee}}, \bibinfo {author} {\bibfnamefont {D.}~\bibnamefont {Meiser}},
  \bibinfo {author} {\bibfnamefont {C.~A.}\ \bibnamefont {Regal}}, \bibinfo
  {author} {\bibfnamefont {K.~W.}\ \bibnamefont {Lehnert}},\ and\ \bibinfo
  {author} {\bibfnamefont {M.~J.}\ \bibnamefont {Holland}},\ }\bibfield
  {title} {\bibinfo {title} {Mechanical resonators for storage and transfer of
  electrical and optical quantum states},\ }\href
  {https://doi.org/10.1103/physreva.87.053818} {\bibfield  {journal} {\bibinfo
  {journal} {Phys. Rev. A}\ }\textbf {\bibinfo {volume} {87}},\ \bibinfo
  {pages} {053818} (\bibinfo {year} {2013})}\BibitemShut {NoStop}%
\bibitem [{\citenamefont {Kurizki}\ \emph {et~al.}(2015)\citenamefont
  {Kurizki}, \citenamefont {Bertet}, \citenamefont {Kubo}, \citenamefont
  {M{\o}lmer}, \citenamefont {Petrosyan}, \citenamefont {Rabl},\ and\
  \citenamefont {Schmiedmayer}}]{Kurizki2015}%
  \BibitemOpen
  \bibfield  {author} {\bibinfo {author} {\bibfnamefont {G.}~\bibnamefont
  {Kurizki}}, \bibinfo {author} {\bibfnamefont {P.}~\bibnamefont {Bertet}},
  \bibinfo {author} {\bibfnamefont {Y.}~\bibnamefont {Kubo}}, \bibinfo {author}
  {\bibfnamefont {K.}~\bibnamefont {M{\o}lmer}}, \bibinfo {author}
  {\bibfnamefont {D.}~\bibnamefont {Petrosyan}}, \bibinfo {author}
  {\bibfnamefont {P.}~\bibnamefont {Rabl}},\ and\ \bibinfo {author}
  {\bibfnamefont {J.}~\bibnamefont {Schmiedmayer}},\ }\bibfield  {title}
  {\bibinfo {title} {Quantum technologies with hybrid systems},\ }\href
  {https://doi.org/10.1073/pnas.1419326112} {\bibfield  {journal} {\bibinfo
  {journal} {Proc. Natl. Acad. Sci.}\ }\textbf {\bibinfo {volume} {112}},\
  \bibinfo {pages} {3866} (\bibinfo {year} {2015})}\BibitemShut {NoStop}%
\bibitem [{\citenamefont {{\v{C}}ernot{\'{\i}}k}\ and\ \citenamefont
  {Hammerer}(2016)}]{Cernotik2016}%
  \BibitemOpen
  \bibfield  {author} {\bibinfo {author} {\bibfnamefont {O.}~\bibnamefont
  {{\v{C}}ernot{\'{\i}}k}}\ and\ \bibinfo {author} {\bibfnamefont
  {K.}~\bibnamefont {Hammerer}},\ }\bibfield  {title} {\bibinfo {title}
  {Measurement-induced long-distance entanglement of superconducting qubits
  using optomechanical transducers},\ }\href
  {https://doi.org/10.1103/physreva.94.012340} {\bibfield  {journal} {\bibinfo
  {journal} {Phys. Rev. A}\ }\textbf {\bibinfo {volume} {94}},\ \bibinfo
  {pages} {012340} (\bibinfo {year} {2016})}\BibitemShut {NoStop}%
\bibitem [{\citenamefont {{\v{C}}ernot{\'{\i}}k}\ \emph
  {et~al.}(2019)\citenamefont {{\v{C}}ernot{\'{\i}}k}, \citenamefont {Genes},\
  and\ \citenamefont {Dantan}}]{Cernotik2019}%
  \BibitemOpen
  \bibfield  {author} {\bibinfo {author} {\bibfnamefont {O.}~\bibnamefont
  {{\v{C}}ernot{\'{\i}}k}}, \bibinfo {author} {\bibfnamefont {C.}~\bibnamefont
  {Genes}},\ and\ \bibinfo {author} {\bibfnamefont {A.}~\bibnamefont
  {Dantan}},\ }\bibfield  {title} {\bibinfo {title} {Interference effects in
  hybrid cavity optomechanics},\ }\href
  {https://doi.org/10.1088/2058-9565/aaf5a6} {\bibfield  {journal} {\bibinfo
  {journal} {Quantum Sci. Technol.}\ }\textbf {\bibinfo {volume} {4}},\
  \bibinfo {pages} {024002} (\bibinfo {year} {2019})}\BibitemShut {NoStop}%
\bibitem [{\citenamefont {Parkins}\ and\ \citenamefont
  {Kimble}(1999)}]{Parkins1999}%
  \BibitemOpen
  \bibfield  {author} {\bibinfo {author} {\bibfnamefont {A.~S.}\ \bibnamefont
  {Parkins}}\ and\ \bibinfo {author} {\bibfnamefont {H.~J.}\ \bibnamefont
  {Kimble}},\ }\bibfield  {title} {\bibinfo {title} {Quantum state transfer
  between motion and light},\ }\href
  {https://doi.org/10.1088/1464-4266/1/4/323} {\bibfield  {journal} {\bibinfo
  {journal} {J. Opt. B: Quantum Semiclass. Opt.}\ }\textbf {\bibinfo {volume}
  {1}},\ \bibinfo {pages} {496} (\bibinfo {year} {1999})}\BibitemShut {NoStop}%
\bibitem [{\citenamefont {Zhang}\ \emph {et~al.}(2003)\citenamefont {Zhang},
  \citenamefont {Peng},\ and\ \citenamefont {Braunstein}}]{Zhang2003}%
  \BibitemOpen
  \bibfield  {author} {\bibinfo {author} {\bibfnamefont {J.}~\bibnamefont
  {Zhang}}, \bibinfo {author} {\bibfnamefont {K.}~\bibnamefont {Peng}},\ and\
  \bibinfo {author} {\bibfnamefont {S.~L.}\ \bibnamefont {Braunstein}},\
  }\bibfield  {title} {\bibinfo {title} {Quantum-state transfer from light to
  macroscopic oscillators},\ }\href
  {https://doi.org/10.1103/physreva.68.013808} {\bibfield  {journal} {\bibinfo
  {journal} {Phys. Rev. A}\ }\textbf {\bibinfo {volume} {68}},\ \bibinfo
  {pages} {013808} (\bibinfo {year} {2003})}\BibitemShut {NoStop}%
\bibitem [{\citenamefont {Safavi-Naeini}\ and\ \citenamefont
  {Painter}(2011)}]{safavi2011proposal}%
  \BibitemOpen
  \bibfield  {author} {\bibinfo {author} {\bibfnamefont {A.~H.}\ \bibnamefont
  {Safavi-Naeini}}\ and\ \bibinfo {author} {\bibfnamefont {O.}~\bibnamefont
  {Painter}},\ }\bibfield  {title} {\bibinfo {title} {Proposal for an
  optomechanical traveling wave phonon--photon translator},\ }\href
  {https://doi.org/10.1088/1367-2630/13/1/013017} {\bibfield  {journal}
  {\bibinfo  {journal} {New J. Phys.}\ }\textbf {\bibinfo {volume} {13}},\
  \bibinfo {pages} {013017} (\bibinfo {year} {2011})}\BibitemShut {NoStop}%
\bibitem [{\citenamefont {Palomaki}\ \emph
  {et~al.}(2013{\natexlab{a}})\citenamefont {Palomaki}, \citenamefont {Harlow},
  \citenamefont {Teufel}, \citenamefont {Simmonds},\ and\ \citenamefont
  {Lehnert}}]{Palomaki2013}%
  \BibitemOpen
  \bibfield  {author} {\bibinfo {author} {\bibfnamefont {T.~A.}\ \bibnamefont
  {Palomaki}}, \bibinfo {author} {\bibfnamefont {J.~W.}\ \bibnamefont
  {Harlow}}, \bibinfo {author} {\bibfnamefont {J.~D.}\ \bibnamefont {Teufel}},
  \bibinfo {author} {\bibfnamefont {R.~W.}\ \bibnamefont {Simmonds}},\ and\
  \bibinfo {author} {\bibfnamefont {K.~W.}\ \bibnamefont {Lehnert}},\
  }\bibfield  {title} {\bibinfo {title} {Coherent state transfer between
  itinerant microwave fields and a mechanical oscillator},\ }\href
  {https://doi.org/10.1038/nature11915} {\bibfield  {journal} {\bibinfo
  {journal} {Nature}\ }\textbf {\bibinfo {volume} {495}},\ \bibinfo {pages}
  {210} (\bibinfo {year} {2013}{\natexlab{a}})}\BibitemShut {NoStop}%
\bibitem [{\citenamefont {{\v{C}}ernot{\'{\i}}k}\ \emph
  {et~al.}(2018)\citenamefont {{\v{C}}ernot{\'{\i}}k}, \citenamefont
  {Mahmoodian},\ and\ \citenamefont {Hammerer}}]{Cernotik2018}%
  \BibitemOpen
  \bibfield  {author} {\bibinfo {author} {\bibfnamefont {O.}~\bibnamefont
  {{\v{C}}ernot{\'{\i}}k}}, \bibinfo {author} {\bibfnamefont {S.}~\bibnamefont
  {Mahmoodian}},\ and\ \bibinfo {author} {\bibfnamefont {K.}~\bibnamefont
  {Hammerer}},\ }\bibfield  {title} {\bibinfo {title} {Spatially adiabatic
  frequency conversion in optoelectromechanical arrays},\ }\href
  {https://doi.org/10.1103/physrevlett.121.110506} {\bibfield  {journal}
  {\bibinfo  {journal} {Phys. Rev. Lett.}\ }\textbf {\bibinfo {volume} {121}},\
  \bibinfo {pages} {110506} (\bibinfo {year} {2018})}\BibitemShut {NoStop}%
\bibitem [{\citenamefont {Wang}\ and\ \citenamefont {Clerk}(2012)}]{wang2012d}%
  \BibitemOpen
  \bibfield  {author} {\bibinfo {author} {\bibfnamefont {Y.-D.}\ \bibnamefont
  {Wang}}\ and\ \bibinfo {author} {\bibfnamefont {A.~A.}\ \bibnamefont
  {Clerk}},\ }\bibfield  {title} {\bibinfo {title} {Using interference for high
  fidelity quantum state transfer in optomechanics},\ }\href
  {https://doi.org/10.1103/PhysRevLett.108.153603} {\bibfield  {journal}
  {\bibinfo  {journal} {Phys. Rev. Lett.}\ }\textbf {\bibinfo {volume} {108}},\
  \bibinfo {pages} {153603} (\bibinfo {year} {2012})}\BibitemShut {NoStop}%
\bibitem [{\citenamefont {Barzanjeh}\ \emph {et~al.}(2012)\citenamefont
  {Barzanjeh}, \citenamefont {Abdi}, \citenamefont {Milburn}, \citenamefont
  {Tombesi},\ and\ \citenamefont {Vitali}}]{Barzanjeh2012}%
  \BibitemOpen
  \bibfield  {author} {\bibinfo {author} {\bibfnamefont {S.}~\bibnamefont
  {Barzanjeh}}, \bibinfo {author} {\bibfnamefont {M.}~\bibnamefont {Abdi}},
  \bibinfo {author} {\bibfnamefont {G.~J.}\ \bibnamefont {Milburn}}, \bibinfo
  {author} {\bibfnamefont {P.}~\bibnamefont {Tombesi}},\ and\ \bibinfo {author}
  {\bibfnamefont {D.}~\bibnamefont {Vitali}},\ }\bibfield  {title} {\bibinfo
  {title} {Reversible optical-to-microwave quantum interface},\ }\href
  {https://doi.org/10.1103/physrevlett.109.130503} {\bibfield  {journal}
  {\bibinfo  {journal} {Phys. Rev. Lett.}\ }\textbf {\bibinfo {volume} {109}},\
  \bibinfo {pages} {130503} (\bibinfo {year} {2012})}\BibitemShut {NoStop}%
\bibitem [{\citenamefont {Tian}(2012)}]{tian2012adiabatic}%
  \BibitemOpen
  \bibfield  {author} {\bibinfo {author} {\bibfnamefont {L.}~\bibnamefont
  {Tian}},\ }\bibfield  {title} {\bibinfo {title} {Adiabatic state conversion
  and pulse transmission in optomechanical systems},\ }\href
  {https://doi.org/10.1103/PhysRevLett.108.153604} {\bibfield  {journal}
  {\bibinfo  {journal} {Phys. Rev. Lett.}\ }\textbf {\bibinfo {volume} {108}},\
  \bibinfo {pages} {153604} (\bibinfo {year} {2012})}\BibitemShut {NoStop}%
\bibitem [{\citenamefont {Schmidt}\ \emph {et~al.}(2012)\citenamefont
  {Schmidt}, \citenamefont {Ludwig},\ and\ \citenamefont
  {Marquardt}}]{Schmidt2012}%
  \BibitemOpen
  \bibfield  {author} {\bibinfo {author} {\bibfnamefont {M.}~\bibnamefont
  {Schmidt}}, \bibinfo {author} {\bibfnamefont {M.}~\bibnamefont {Ludwig}},\
  and\ \bibinfo {author} {\bibfnamefont {F.}~\bibnamefont {Marquardt}},\
  }\bibfield  {title} {\bibinfo {title} {Optomechanical circuits for
  nanomechanical continuous variable quantum state processing},\ }\href
  {https://doi.org/10.1088/1367-2630/14/12/125005} {\bibfield  {journal}
  {\bibinfo  {journal} {New J. Phys.}\ }\textbf {\bibinfo {volume} {14}},\
  \bibinfo {pages} {125005} (\bibinfo {year} {2012})}\BibitemShut {NoStop}%
\bibitem [{\citenamefont {Sete}\ and\ \citenamefont {Eleuch}(2015)}]{Sete2015}%
  \BibitemOpen
  \bibfield  {author} {\bibinfo {author} {\bibfnamefont {E.~A.}\ \bibnamefont
  {Sete}}\ and\ \bibinfo {author} {\bibfnamefont {H.}~\bibnamefont {Eleuch}},\
  }\bibfield  {title} {\bibinfo {title} {High-efficiency quantum state transfer
  and quantum memory using a mechanical oscillator},\ }\href
  {https://doi.org/10.1103/PhysRevA.91.032309} {\bibfield  {journal} {\bibinfo
  {journal} {Phys. Rev. A}\ }\textbf {\bibinfo {volume} {91}},\ \bibinfo
  {pages} {032309} (\bibinfo {year} {2015})}\BibitemShut {NoStop}%
\bibitem [{\citenamefont {Hill}\ \emph {et~al.}(2012)\citenamefont {Hill},
  \citenamefont {Safavi-Naeini}, \citenamefont {Chan},\ and\ \citenamefont
  {Painter}}]{hill2012coherent}%
  \BibitemOpen
  \bibfield  {author} {\bibinfo {author} {\bibfnamefont {J.~T.}\ \bibnamefont
  {Hill}}, \bibinfo {author} {\bibfnamefont {A.~H.}\ \bibnamefont
  {Safavi-Naeini}}, \bibinfo {author} {\bibfnamefont {J.}~\bibnamefont
  {Chan}},\ and\ \bibinfo {author} {\bibfnamefont {O.}~\bibnamefont
  {Painter}},\ }\bibfield  {title} {\bibinfo {title} {Coherent optical
  wavelength conversion via cavity optomechanics},\ }\href
  {https://doi.org/10.1038/ncomms2201} {\bibfield  {journal} {\bibinfo
  {journal} {Nat. Commun.}\ }\textbf {\bibinfo {volume} {3}},\ \bibinfo {pages}
  {1196} (\bibinfo {year} {2012})}\BibitemShut {NoStop}%
\bibitem [{\citenamefont {Bochmann}\ \emph {et~al.}(2013)\citenamefont
  {Bochmann}, \citenamefont {Vainsencher}, \citenamefont {Awschalom},\ and\
  \citenamefont {Cleland}}]{bochmann2013nanomechanical}%
  \BibitemOpen
  \bibfield  {author} {\bibinfo {author} {\bibfnamefont {J.}~\bibnamefont
  {Bochmann}}, \bibinfo {author} {\bibfnamefont {A.}~\bibnamefont
  {Vainsencher}}, \bibinfo {author} {\bibfnamefont {D.~D.}\ \bibnamefont
  {Awschalom}},\ and\ \bibinfo {author} {\bibfnamefont {A.~N.}\ \bibnamefont
  {Cleland}},\ }\bibfield  {title} {\bibinfo {title} {Nanomechanical coupling
  between microwave and optical photons},\ }\href
  {https://doi.org/10.1038/nphys2748} {\bibfield  {journal} {\bibinfo
  {journal} {Nat. Phys.}\ }\textbf {\bibinfo {volume} {9}},\ \bibinfo {pages}
  {712} (\bibinfo {year} {2013})}\BibitemShut {NoStop}%
\bibitem [{\citenamefont {Andrews}\ \emph {et~al.}(2014)\citenamefont
  {Andrews}, \citenamefont {Peterson}, \citenamefont {Purdy}, \citenamefont
  {Cicak}, \citenamefont {Simmonds}, \citenamefont {Regal},\ and\ \citenamefont
  {Lehnert}}]{andrews2014bidirectional}%
  \BibitemOpen
  \bibfield  {author} {\bibinfo {author} {\bibfnamefont {R.~W.}\ \bibnamefont
  {Andrews}}, \bibinfo {author} {\bibfnamefont {R.~W.}\ \bibnamefont
  {Peterson}}, \bibinfo {author} {\bibfnamefont {T.~P.}\ \bibnamefont {Purdy}},
  \bibinfo {author} {\bibfnamefont {K.}~\bibnamefont {Cicak}}, \bibinfo
  {author} {\bibfnamefont {R.~W.}\ \bibnamefont {Simmonds}}, \bibinfo {author}
  {\bibfnamefont {C.~A.}\ \bibnamefont {Regal}},\ and\ \bibinfo {author}
  {\bibfnamefont {K.~W.}\ \bibnamefont {Lehnert}},\ }\bibfield  {title}
  {\bibinfo {title} {Bidirectional and efficient conversion between microwave
  and optical light},\ }\href {https://doi.org/10.1038/nphys2911} {\bibfield
  {journal} {\bibinfo  {journal} {Nat. Phys.}\ }\textbf {\bibinfo {volume}
  {10}},\ \bibinfo {pages} {321} (\bibinfo {year} {2014})}\BibitemShut
  {NoStop}%
\bibitem [{\citenamefont {Balram}\ \emph {et~al.}(2016)\citenamefont {Balram},
  \citenamefont {Davan{\c{c}}o}, \citenamefont {Song},\ and\ \citenamefont
  {Srinivasan}}]{balram2016coherent}%
  \BibitemOpen
  \bibfield  {author} {\bibinfo {author} {\bibfnamefont {K.~C.}\ \bibnamefont
  {Balram}}, \bibinfo {author} {\bibfnamefont {M.~I.}\ \bibnamefont
  {Davan{\c{c}}o}}, \bibinfo {author} {\bibfnamefont {J.~D.}\ \bibnamefont
  {Song}},\ and\ \bibinfo {author} {\bibfnamefont {K.}~\bibnamefont
  {Srinivasan}},\ }\bibfield  {title} {\bibinfo {title} {Coherent coupling
  between radiofrequency, optical and acoustic waves in piezo-optomechanical
  circuits},\ }\href {https://doi.org/10.1038/nphoton.2016.46} {\bibfield
  {journal} {\bibinfo  {journal} {Nat. Photonics}\ }\textbf {\bibinfo {volume}
  {10}},\ \bibinfo {pages} {346} (\bibinfo {year} {2016})}\BibitemShut
  {NoStop}%
\bibitem [{\citenamefont {Habraken}\ \emph {et~al.}(2012)\citenamefont
  {Habraken}, \citenamefont {Stannigel}, \citenamefont {Lukin}, \citenamefont
  {Zoller},\ and\ \citenamefont {Rabl}}]{habraken2012continuous}%
  \BibitemOpen
  \bibfield  {author} {\bibinfo {author} {\bibfnamefont {S.}~\bibnamefont
  {Habraken}}, \bibinfo {author} {\bibfnamefont {K.}~\bibnamefont {Stannigel}},
  \bibinfo {author} {\bibfnamefont {M.~D.}\ \bibnamefont {Lukin}}, \bibinfo
  {author} {\bibfnamefont {P.}~\bibnamefont {Zoller}},\ and\ \bibinfo {author}
  {\bibfnamefont {P.}~\bibnamefont {Rabl}},\ }\bibfield  {title} {\bibinfo
  {title} {Continuous mode cooling and phonon routers for phononic quantum
  networks},\ }\href {https://doi.org/10.1088/1367-2630/14/11/115004}
  {\bibfield  {journal} {\bibinfo  {journal} {New J. Phys.}\ }\textbf {\bibinfo
  {volume} {14}},\ \bibinfo {pages} {115004} (\bibinfo {year}
  {2012})}\BibitemShut {NoStop}%
\bibitem [{\citenamefont {Xuereb}\ \emph {et~al.}(2012)\citenamefont {Xuereb},
  \citenamefont {Genes},\ and\ \citenamefont {Dantan}}]{Xuereb2012}%
  \BibitemOpen
  \bibfield  {author} {\bibinfo {author} {\bibfnamefont {A.}~\bibnamefont
  {Xuereb}}, \bibinfo {author} {\bibfnamefont {C.}~\bibnamefont {Genes}},\ and\
  \bibinfo {author} {\bibfnamefont {A.}~\bibnamefont {Dantan}},\ }\bibfield
  {title} {\bibinfo {title} {Strong coupling and long-range collective
  interactions in optomechanical arrays},\ }\href
  {https://doi.org/10.1103/physrevlett.109.223601} {\bibfield  {journal}
  {\bibinfo  {journal} {Phys. Rev. Lett.}\ }\textbf {\bibinfo {volume} {109}},\
  \bibinfo {pages} {223601} (\bibinfo {year} {2012})}\BibitemShut {NoStop}%
\bibitem [{\citenamefont {Abdi}\ \emph {et~al.}(2012)\citenamefont {Abdi},
  \citenamefont {Pirandola}, \citenamefont {Tombesi},\ and\ \citenamefont
  {Vitali}}]{Abdi2012}%
  \BibitemOpen
  \bibfield  {author} {\bibinfo {author} {\bibfnamefont {M.}~\bibnamefont
  {Abdi}}, \bibinfo {author} {\bibfnamefont {S.}~\bibnamefont {Pirandola}},
  \bibinfo {author} {\bibfnamefont {P.}~\bibnamefont {Tombesi}},\ and\ \bibinfo
  {author} {\bibfnamefont {D.}~\bibnamefont {Vitali}},\ }\bibfield  {title}
  {\bibinfo {title} {Entanglement swapping with local certification:
  Application to remote micromechanical resonators},\ }\href
  {https://doi.org/10.1103/physrevlett.109.143601} {\bibfield  {journal}
  {\bibinfo  {journal} {Phys. Rev. Lett.}\ }\textbf {\bibinfo {volume} {109}},\
  \bibinfo {pages} {143601} (\bibinfo {year} {2012})}\BibitemShut {NoStop}%
\bibitem [{\citenamefont {Singh}\ \emph {et~al.}(2012)\citenamefont {Singh},
  \citenamefont {Jing}, \citenamefont {Wright},\ and\ \citenamefont
  {Meystre}}]{Singh2012}%
  \BibitemOpen
  \bibfield  {author} {\bibinfo {author} {\bibfnamefont {S.}~\bibnamefont
  {Singh}}, \bibinfo {author} {\bibfnamefont {H.}~\bibnamefont {Jing}},
  \bibinfo {author} {\bibfnamefont {E.~M.}\ \bibnamefont {Wright}},\ and\
  \bibinfo {author} {\bibfnamefont {P.}~\bibnamefont {Meystre}},\ }\bibfield
  {title} {\bibinfo {title} {Quantum-state transfer between a bose-einstein
  condensate and an optomechanical mirror},\ }\href
  {https://doi.org/10.1103/physreva.86.021801} {\bibfield  {journal} {\bibinfo
  {journal} {Phys. Rev. A}\ }\textbf {\bibinfo {volume} {86}},\ \bibinfo
  {pages} {021801} (\bibinfo {year} {2012})}\BibitemShut {NoStop}%
\bibitem [{\citenamefont {Shkarin}\ \emph {et~al.}(2014)\citenamefont
  {Shkarin}, \citenamefont {Flowers-Jacobs}, \citenamefont {Hoch},
  \citenamefont {Kashkanova}, \citenamefont {Deutsch}, \citenamefont
  {Reichel},\ and\ \citenamefont {Harris}}]{shkarin2014optically}%
  \BibitemOpen
  \bibfield  {author} {\bibinfo {author} {\bibfnamefont {A.~B.}\ \bibnamefont
  {Shkarin}}, \bibinfo {author} {\bibfnamefont {N.~E.}\ \bibnamefont
  {Flowers-Jacobs}}, \bibinfo {author} {\bibfnamefont {S.~W.}\ \bibnamefont
  {Hoch}}, \bibinfo {author} {\bibfnamefont {A.~D.}\ \bibnamefont
  {Kashkanova}}, \bibinfo {author} {\bibfnamefont {C.}~\bibnamefont {Deutsch}},
  \bibinfo {author} {\bibfnamefont {J.}~\bibnamefont {Reichel}},\ and\ \bibinfo
  {author} {\bibfnamefont {J.~G.~E.}\ \bibnamefont {Harris}},\ }\bibfield
  {title} {\bibinfo {title} {Optically mediated hybridization between two
  mechanical modes},\ }\href {https://doi.org/10.1103/PhysRevLett.112.013602}
  {\bibfield  {journal} {\bibinfo  {journal} {Phys. Rev. Lett.}\ }\textbf
  {\bibinfo {volume} {112}},\ \bibinfo {pages} {013602} (\bibinfo {year}
  {2014})}\BibitemShut {NoStop}%
\bibitem [{\citenamefont {Abdi}\ and\ \citenamefont
  {Hartmann}(2015)}]{Abdi2015}%
  \BibitemOpen
  \bibfield  {author} {\bibinfo {author} {\bibfnamefont {M.}~\bibnamefont
  {Abdi}}\ and\ \bibinfo {author} {\bibfnamefont {M.~J.}\ \bibnamefont
  {Hartmann}},\ }\bibfield  {title} {\bibinfo {title} {Entangling the motion of
  two optically trapped objects via time-modulated driving fields},\ }\href
  {https://doi.org/10.1088/1367-2630/17/1/013056} {\bibfield  {journal}
  {\bibinfo  {journal} {New J. Phys.}\ }\textbf {\bibinfo {volume} {17}},\
  \bibinfo {pages} {013056} (\bibinfo {year} {2015})}\BibitemShut {NoStop}%
\bibitem [{\citenamefont {Weaver}\ \emph {et~al.}(2017)\citenamefont {Weaver},
  \citenamefont {Buters}, \citenamefont {Luna}, \citenamefont {Eerkens},
  \citenamefont {Heeck}, \citenamefont {de~Man},\ and\ \citenamefont
  {Bouwmeester}}]{Weaver2017}%
  \BibitemOpen
  \bibfield  {author} {\bibinfo {author} {\bibfnamefont {M.~J.}\ \bibnamefont
  {Weaver}}, \bibinfo {author} {\bibfnamefont {F.}~\bibnamefont {Buters}},
  \bibinfo {author} {\bibfnamefont {F.}~\bibnamefont {Luna}}, \bibinfo {author}
  {\bibfnamefont {H.}~\bibnamefont {Eerkens}}, \bibinfo {author} {\bibfnamefont
  {K.}~\bibnamefont {Heeck}}, \bibinfo {author} {\bibfnamefont
  {S.}~\bibnamefont {de~Man}},\ and\ \bibinfo {author} {\bibfnamefont
  {D.}~\bibnamefont {Bouwmeester}},\ }\bibfield  {title} {\bibinfo {title}
  {Coherent optomechanical state transfer between disparate mechanical
  resonators},\ }\bibfield  {journal} {\bibinfo  {journal} {Nat. Commun.}\
  }\textbf {\bibinfo {volume} {8}},\ \href
  {https://doi.org/10.1038/s41467-017-00968-9} {10.1038/s41467-017-00968-9}
  (\bibinfo {year} {2017})\BibitemShut {NoStop}%
\bibitem [{\citenamefont {Zhang}\ \emph {et~al.}(2020)\citenamefont {Zhang},
  \citenamefont {Song}, \citenamefont {Luo}, \citenamefont {Su}, \citenamefont
  {Wang}, \citenamefont {Cao}, \citenamefont {Li}, \citenamefont {Xiao},
  \citenamefont {Guo}, \citenamefont {Tian}, \citenamefont {Deng},\ and\
  \citenamefont {Guo}}]{Zhang2020}%
  \BibitemOpen
  \bibfield  {author} {\bibinfo {author} {\bibfnamefont {Z.-Z.}\ \bibnamefont
  {Zhang}}, \bibinfo {author} {\bibfnamefont {X.-X.}\ \bibnamefont {Song}},
  \bibinfo {author} {\bibfnamefont {G.}~\bibnamefont {Luo}}, \bibinfo {author}
  {\bibfnamefont {Z.-J.}\ \bibnamefont {Su}}, \bibinfo {author} {\bibfnamefont
  {K.-L.}\ \bibnamefont {Wang}}, \bibinfo {author} {\bibfnamefont
  {G.}~\bibnamefont {Cao}}, \bibinfo {author} {\bibfnamefont {H.-O.}\
  \bibnamefont {Li}}, \bibinfo {author} {\bibfnamefont {M.}~\bibnamefont
  {Xiao}}, \bibinfo {author} {\bibfnamefont {G.-C.}\ \bibnamefont {Guo}},
  \bibinfo {author} {\bibfnamefont {L.}~\bibnamefont {Tian}}, \bibinfo {author}
  {\bibfnamefont {G.-W.}\ \bibnamefont {Deng}},\ and\ \bibinfo {author}
  {\bibfnamefont {G.-P.}\ \bibnamefont {Guo}},\ }\bibfield  {title} {\bibinfo
  {title} {Coherent phonon dynamics in spatially separated graphene mechanical
  resonators},\ }\href {https://doi.org/10.1073/pnas.1916978117} {\bibfield
  {journal} {\bibinfo  {journal} {Proc. Natl. Acad. Sci.}\ }\textbf {\bibinfo
  {volume} {117}},\ \bibinfo {pages} {5582} (\bibinfo {year}
  {2020})}\BibitemShut {NoStop}%
\bibitem [{\citenamefont {Abdi}(2021)}]{Abdi2021}%
  \BibitemOpen
  \bibfield  {author} {\bibinfo {author} {\bibfnamefont {M.}~\bibnamefont
  {Abdi}},\ }\bibfield  {title} {\bibinfo {title} {Continuous-variable
  multipartite vibrational entanglement},\ }\href
  {https://doi.org/10.1103/physreva.103.043520} {\bibfield  {journal} {\bibinfo
   {journal} {Phys. Rev. A}\ }\textbf {\bibinfo {volume} {103}},\ \bibinfo
  {pages} {043520} (\bibinfo {year} {2021})}\BibitemShut {NoStop}%
\bibitem [{\citenamefont {Felicetti}\ \emph {et~al.}(2017)\citenamefont
  {Felicetti}, \citenamefont {Fedortchenko}, \citenamefont {Rossi~Jr},
  \citenamefont {Ducci}, \citenamefont {Favero}, \citenamefont {Coudreau},\
  and\ \citenamefont {Milman}}]{felicetti2017quantum}%
  \BibitemOpen
  \bibfield  {author} {\bibinfo {author} {\bibfnamefont {S.}~\bibnamefont
  {Felicetti}}, \bibinfo {author} {\bibfnamefont {S.}~\bibnamefont
  {Fedortchenko}}, \bibinfo {author} {\bibfnamefont {R.}~\bibnamefont
  {Rossi~Jr}}, \bibinfo {author} {\bibfnamefont {S.}~\bibnamefont {Ducci}},
  \bibinfo {author} {\bibfnamefont {I.}~\bibnamefont {Favero}}, \bibinfo
  {author} {\bibfnamefont {T.}~\bibnamefont {Coudreau}},\ and\ \bibinfo
  {author} {\bibfnamefont {P.}~\bibnamefont {Milman}},\ }\bibfield  {title}
  {\bibinfo {title} {Quantum communication between remote mechanical
  resonators},\ }\href {https://doi.org/10.1103/PhysRevA.95.022322} {\bibfield
  {journal} {\bibinfo  {journal} {Phys. Rev. A}\ }\textbf {\bibinfo {volume}
  {95}},\ \bibinfo {pages} {022322} (\bibinfo {year} {2017})}\BibitemShut
  {NoStop}%
\bibitem [{\citenamefont {Pautrel}\ \emph {et~al.}(2020)\citenamefont
  {Pautrel}, \citenamefont {Denis}, \citenamefont {Bon}, \citenamefont
  {Borne},\ and\ \citenamefont {Favero}}]{Pautrel2020}%
  \BibitemOpen
  \bibfield  {author} {\bibinfo {author} {\bibfnamefont {S.}~\bibnamefont
  {Pautrel}}, \bibinfo {author} {\bibfnamefont {Z.}~\bibnamefont {Denis}},
  \bibinfo {author} {\bibfnamefont {J.}~\bibnamefont {Bon}}, \bibinfo {author}
  {\bibfnamefont {A.}~\bibnamefont {Borne}},\ and\ \bibinfo {author}
  {\bibfnamefont {I.}~\bibnamefont {Favero}},\ }\bibfield  {title} {\bibinfo
  {title} {Optomechanical discrete-variable quantum teleportation scheme},\
  }\href {https://doi.org/10.1103/physreva.101.063820} {\bibfield  {journal}
  {\bibinfo  {journal} {Phys. Rev. A}\ }\textbf {\bibinfo {volume} {101}},\
  \bibinfo {pages} {063820} (\bibinfo {year} {2020})}\BibitemShut {NoStop}%
\bibitem [{\citenamefont {Li}\ \emph {et~al.}(2020)\citenamefont {Li},
  \citenamefont {Wallucks}, \citenamefont {Benevides}, \citenamefont {Fiaschi},
  \citenamefont {Hensen}, \citenamefont {Alegre},\ and\ \citenamefont
  {Gröblacher}}]{Li2020}%
  \BibitemOpen
  \bibfield  {author} {\bibinfo {author} {\bibfnamefont {J.}~\bibnamefont
  {Li}}, \bibinfo {author} {\bibfnamefont {A.}~\bibnamefont {Wallucks}},
  \bibinfo {author} {\bibfnamefont {R.}~\bibnamefont {Benevides}}, \bibinfo
  {author} {\bibfnamefont {N.}~\bibnamefont {Fiaschi}}, \bibinfo {author}
  {\bibfnamefont {B.}~\bibnamefont {Hensen}}, \bibinfo {author} {\bibfnamefont
  {T.~P.~M.}\ \bibnamefont {Alegre}},\ and\ \bibinfo {author} {\bibfnamefont
  {S.}~\bibnamefont {Gröblacher}},\ }\bibfield  {title} {\bibinfo {title}
  {Proposal for optomechanical quantum teleportation},\ }\href
  {https://doi.org/10.1103/physreva.102.032402} {\bibfield  {journal} {\bibinfo
   {journal} {Phys. Rev. A}\ }\textbf {\bibinfo {volume} {102}},\ \bibinfo
  {pages} {032402} (\bibinfo {year} {2020})}\BibitemShut {NoStop}%
\bibitem [{\citenamefont {Berry}(2009)}]{berry2009transitionless}%
  \BibitemOpen
  \bibfield  {author} {\bibinfo {author} {\bibfnamefont {M.~V.}\ \bibnamefont
  {Berry}},\ }\bibfield  {title} {\bibinfo {title} {Transitionless quantum
  driving},\ }\href {https://doi.org/10.1088/1751-8113/42/36/365303} {\bibfield
   {journal} {\bibinfo  {journal} {J. Phys. A}\ }\textbf {\bibinfo {volume}
  {42}},\ \bibinfo {pages} {365303} (\bibinfo {year} {2009})}\BibitemShut
  {NoStop}%
\bibitem [{\citenamefont {Chen}\ \emph {et~al.}(2010)\citenamefont {Chen},
  \citenamefont {Lizuain}, \citenamefont {Ruschhaupt}, \citenamefont
  {Gu{\'e}ry-Odelin},\ and\ \citenamefont {Muga}}]{chen2010shortcut}%
  \BibitemOpen
  \bibfield  {author} {\bibinfo {author} {\bibfnamefont {X.}~\bibnamefont
  {Chen}}, \bibinfo {author} {\bibfnamefont {I.}~\bibnamefont {Lizuain}},
  \bibinfo {author} {\bibfnamefont {A.}~\bibnamefont {Ruschhaupt}}, \bibinfo
  {author} {\bibfnamefont {D.}~\bibnamefont {Gu{\'e}ry-Odelin}},\ and\ \bibinfo
  {author} {\bibfnamefont {J.~G.}\ \bibnamefont {Muga}},\ }\bibfield  {title}
  {\bibinfo {title} {Shortcut to adiabatic passage in two-and three-level
  atoms},\ }\href {https://doi.org/10.1103/PhysRevLett.105.123003} {\bibfield
  {journal} {\bibinfo  {journal} {Phys. Rev. Lett.}\ }\textbf {\bibinfo
  {volume} {105}},\ \bibinfo {pages} {123003} (\bibinfo {year}
  {2010})}\BibitemShut {NoStop}%
\bibitem [{\citenamefont {Weedbrook}\ \emph {et~al.}(2012)\citenamefont
  {Weedbrook}, \citenamefont {Pirandola}, \citenamefont
  {Garc{\'{\i}}a-Patr{\'{o}}n}, \citenamefont {Cerf}, \citenamefont {Ralph},
  \citenamefont {Shapiro},\ and\ \citenamefont {Lloyd}}]{Weedbrook2012}%
  \BibitemOpen
  \bibfield  {author} {\bibinfo {author} {\bibfnamefont {C.}~\bibnamefont
  {Weedbrook}}, \bibinfo {author} {\bibfnamefont {S.}~\bibnamefont
  {Pirandola}}, \bibinfo {author} {\bibfnamefont {R.}~\bibnamefont
  {Garc{\'{\i}}a-Patr{\'{o}}n}}, \bibinfo {author} {\bibfnamefont {N.~J.}\
  \bibnamefont {Cerf}}, \bibinfo {author} {\bibfnamefont {T.~C.}\ \bibnamefont
  {Ralph}}, \bibinfo {author} {\bibfnamefont {J.~H.}\ \bibnamefont {Shapiro}},\
  and\ \bibinfo {author} {\bibfnamefont {S.}~\bibnamefont {Lloyd}},\ }\bibfield
   {title} {\bibinfo {title} {Gaussian quantum information},\ }\href
  {https://doi.org/10.1103/revmodphys.84.621} {\bibfield  {journal} {\bibinfo
  {journal} {Rev. Mod. Phys.}\ }\textbf {\bibinfo {volume} {84}},\ \bibinfo
  {pages} {621} (\bibinfo {year} {2012})}\BibitemShut {NoStop}%
\bibitem [{\citenamefont {Aspelmeyer}\ \emph {et~al.}(2014)\citenamefont
  {Aspelmeyer}, \citenamefont {Kippenberg},\ and\ \citenamefont
  {Marquardt}}]{aspelmeyer2014cavity}%
  \BibitemOpen
  \bibfield  {author} {\bibinfo {author} {\bibfnamefont {M.}~\bibnamefont
  {Aspelmeyer}}, \bibinfo {author} {\bibfnamefont {T.~J.}\ \bibnamefont
  {Kippenberg}},\ and\ \bibinfo {author} {\bibfnamefont {F.}~\bibnamefont
  {Marquardt}},\ }\bibfield  {title} {\bibinfo {title} {Cavity optomechanics},\
  }\href {https://doi.org/10.1103/RevModPhys.86.1391} {\bibfield  {journal}
  {\bibinfo  {journal} {Rev. Mod. Phys.}\ }\textbf {\bibinfo {volume} {86}},\
  \bibinfo {pages} {1391} (\bibinfo {year} {2014})}\BibitemShut {NoStop}%
\bibitem [{\citenamefont {Abdi}\ \emph {et~al.}(2011)\citenamefont {Abdi},
  \citenamefont {Barzanjeh}, \citenamefont {Tombesi},\ and\ \citenamefont
  {Vitali}}]{Abdi2011}%
  \BibitemOpen
  \bibfield  {author} {\bibinfo {author} {\bibfnamefont {M.}~\bibnamefont
  {Abdi}}, \bibinfo {author} {\bibfnamefont {S.}~\bibnamefont {Barzanjeh}},
  \bibinfo {author} {\bibfnamefont {P.}~\bibnamefont {Tombesi}},\ and\ \bibinfo
  {author} {\bibfnamefont {D.}~\bibnamefont {Vitali}},\ }\bibfield  {title}
  {\bibinfo {title} {Effect of phase noise on the generation of stationary
  entanglement in cavity optomechanics},\ }\href
  {https://doi.org/10.1103/physreva.84.032325} {\bibfield  {journal} {\bibinfo
  {journal} {Phys. Rev. A}\ }\textbf {\bibinfo {volume} {84}},\ \bibinfo
  {pages} {032325} (\bibinfo {year} {2011})}\BibitemShut {NoStop}%
\bibitem [{\citenamefont {Palomaki}\ \emph
  {et~al.}(2013{\natexlab{b}})\citenamefont {Palomaki}, \citenamefont {Teufel},
  \citenamefont {Simmonds},\ and\ \citenamefont {Lehnert}}]{Palomaki2013a}%
  \BibitemOpen
  \bibfield  {author} {\bibinfo {author} {\bibfnamefont {T.~A.}\ \bibnamefont
  {Palomaki}}, \bibinfo {author} {\bibfnamefont {J.~D.}\ \bibnamefont
  {Teufel}}, \bibinfo {author} {\bibfnamefont {R.~W.}\ \bibnamefont
  {Simmonds}},\ and\ \bibinfo {author} {\bibfnamefont {K.~W.}\ \bibnamefont
  {Lehnert}},\ }\bibfield  {title} {\bibinfo {title} {Entangling mechanical
  motion with microwave fields},\ }\href
  {https://doi.org/10.1126/science.1244563} {\bibfield  {journal} {\bibinfo
  {journal} {Science}\ }\textbf {\bibinfo {volume} {342}},\ \bibinfo {pages}
  {710} (\bibinfo {year} {2013}{\natexlab{b}})}\BibitemShut {NoStop}%
\bibitem [{\citenamefont {Patel}\ \emph {et~al.}(2018)\citenamefont {Patel},
  \citenamefont {Wang}, \citenamefont {Jiang}, \citenamefont {Sarabalis},
  \citenamefont {Hill},\ and\ \citenamefont {Safavi-Naeini}}]{patel2018single}%
  \BibitemOpen
  \bibfield  {author} {\bibinfo {author} {\bibfnamefont {R.~N.}\ \bibnamefont
  {Patel}}, \bibinfo {author} {\bibfnamefont {Z.}~\bibnamefont {Wang}},
  \bibinfo {author} {\bibfnamefont {W.}~\bibnamefont {Jiang}}, \bibinfo
  {author} {\bibfnamefont {C.~J.}\ \bibnamefont {Sarabalis}}, \bibinfo {author}
  {\bibfnamefont {J.~T.}\ \bibnamefont {Hill}},\ and\ \bibinfo {author}
  {\bibfnamefont {A.~H.}\ \bibnamefont {Safavi-Naeini}},\ }\bibfield  {title}
  {\bibinfo {title} {Single-mode phononic wire},\ }\href
  {https://doi.org/10.1103/PhysRevLett.121.040501} {\bibfield  {journal}
  {\bibinfo  {journal} {Phys. Rev. Lett.}\ }\textbf {\bibinfo {volume} {121}},\
  \bibinfo {pages} {040501} (\bibinfo {year} {2018})}\BibitemShut {NoStop}%
\bibitem [{\citenamefont {Stannigel}\ \emph {et~al.}(2011)\citenamefont
  {Stannigel}, \citenamefont {Rabl}, \citenamefont {S{\o}rensen}, \citenamefont
  {Lukin},\ and\ \citenamefont {Zoller}}]{stannigel2011optomechanical}%
  \BibitemOpen
  \bibfield  {author} {\bibinfo {author} {\bibfnamefont {K.}~\bibnamefont
  {Stannigel}}, \bibinfo {author} {\bibfnamefont {P.}~\bibnamefont {Rabl}},
  \bibinfo {author} {\bibfnamefont {A.~S.}\ \bibnamefont {S{\o}rensen}},
  \bibinfo {author} {\bibfnamefont {M.~D.}\ \bibnamefont {Lukin}},\ and\
  \bibinfo {author} {\bibfnamefont {P.}~\bibnamefont {Zoller}},\ }\bibfield
  {title} {\bibinfo {title} {Optomechanical transducers for quantum-information
  processing},\ }\href {https://link.aps.org/doi/10.1103/PhysRevA.84.042341.}
  {\bibfield  {journal} {\bibinfo  {journal} {Phys. Rev. A}\ }\textbf {\bibinfo
  {volume} {84}},\ \bibinfo {pages} {042341} (\bibinfo {year}
  {2011})}\BibitemShut {NoStop}%
\bibitem [{\citenamefont {Derezi{\'{n}}ski}(2017)}]{Derezinski2017}%
  \BibitemOpen
  \bibfield  {author} {\bibinfo {author} {\bibfnamefont {J.}~\bibnamefont
  {Derezi{\'{n}}ski}},\ }\bibfield  {title} {\bibinfo {title} {Bosonic
  quadratic hamiltonians},\ }\href {https://doi.org/10.1063/1.5017931}
  {\bibfield  {journal} {\bibinfo  {journal} {J. Math. Phys.}\ }\textbf
  {\bibinfo {volume} {58}},\ \bibinfo {pages} {121101} (\bibinfo {year}
  {2017})}\BibitemShut {NoStop}%
\bibitem [{\citenamefont {Banchi}\ \emph {et~al.}(2015)\citenamefont {Banchi},
  \citenamefont {Braunstein},\ and\ \citenamefont {Pirandola}}]{Banchi2015}%
  \BibitemOpen
  \bibfield  {author} {\bibinfo {author} {\bibfnamefont {L.}~\bibnamefont
  {Banchi}}, \bibinfo {author} {\bibfnamefont {S.~L.}\ \bibnamefont
  {Braunstein}},\ and\ \bibinfo {author} {\bibfnamefont {S.}~\bibnamefont
  {Pirandola}},\ }\bibfield  {title} {\bibinfo {title} {Quantum fidelity for
  arbitrary gaussian states},\ }\href
  {https://doi.org/10.1103/PhysRevLett.115.260501} {\bibfield  {journal}
  {\bibinfo  {journal} {Phys. Rev. Lett.}\ }\textbf {\bibinfo {volume} {115}},\
  \bibinfo {pages} {260501} (\bibinfo {year} {2015})}\BibitemShut {NoStop}%
\bibitem [{\citenamefont {Gu{\'{e}}ry-Odelin}\ \emph
  {et~al.}(2019)\citenamefont {Gu{\'{e}}ry-Odelin}, \citenamefont {Ruschhaupt},
  \citenamefont {Kiely}, \citenamefont {Torrontegui}, \citenamefont
  {Mart{\'{\i}}nez-Garaot},\ and\ \citenamefont {Muga}}]{GueryOdelin2019}%
  \BibitemOpen
  \bibfield  {author} {\bibinfo {author} {\bibfnamefont {D.}~\bibnamefont
  {Gu{\'{e}}ry-Odelin}}, \bibinfo {author} {\bibfnamefont {A.}~\bibnamefont
  {Ruschhaupt}}, \bibinfo {author} {\bibfnamefont {A.}~\bibnamefont {Kiely}},
  \bibinfo {author} {\bibfnamefont {E.}~\bibnamefont {Torrontegui}}, \bibinfo
  {author} {\bibfnamefont {S.}~\bibnamefont {Mart{\'{\i}}nez-Garaot}},\ and\
  \bibinfo {author} {\bibfnamefont {J.}~\bibnamefont {Muga}},\ }\bibfield
  {title} {\bibinfo {title} {Shortcuts to adiabaticity: Concepts, methods, and
  applications},\ }\href {https://doi.org/10.1103/revmodphys.91.045001}
  {\bibfield  {journal} {\bibinfo  {journal} {Rev. Mod. Phys.}\ }\textbf
  {\bibinfo {volume} {91}},\ \bibinfo {pages} {045001} (\bibinfo {year}
  {2019})}\BibitemShut {NoStop}%
\bibitem [{\citenamefont {Zhang}\ \emph {et~al.}(2019)\citenamefont {Zhang},
  \citenamefont {Li}, \citenamefont {Yan},\ and\ \citenamefont
  {Xia}}]{Zhang2019}%
  \BibitemOpen
  \bibfield  {author} {\bibinfo {author} {\bibfnamefont {F.-Y.}\ \bibnamefont
  {Zhang}}, \bibinfo {author} {\bibfnamefont {W.-L.}\ \bibnamefont {Li}},
  \bibinfo {author} {\bibfnamefont {W.-B.}\ \bibnamefont {Yan}},\ and\ \bibinfo
  {author} {\bibfnamefont {Y.}~\bibnamefont {Xia}},\ }\bibfield  {title}
  {\bibinfo {title} {Speeding up adiabatic state conversion in optomechanical
  systems},\ }\href {https://doi.org/10.1088/1361-6455/ab08d8} {\bibfield
  {journal} {\bibinfo  {journal} {J. Phys. B: At. Mol. Opt. Phys.}\ }\textbf
  {\bibinfo {volume} {52}},\ \bibinfo {pages} {115501} (\bibinfo {year}
  {2019})}\BibitemShut {NoStop}%
\bibitem [{\citenamefont {Park}\ and\ \citenamefont {Wang}(2009)}]{Park2009}%
  \BibitemOpen
  \bibfield  {author} {\bibinfo {author} {\bibfnamefont {Y.-S.}\ \bibnamefont
  {Park}}\ and\ \bibinfo {author} {\bibfnamefont {H.}~\bibnamefont {Wang}},\
  }\bibfield  {title} {\bibinfo {title} {Resolved-sideband and cryogenic
  cooling of an optomechanical resonator},\ }\href
  {https://doi.org/10.1038/nphys1303} {\bibfield  {journal} {\bibinfo
  {journal} {Nat. Phys.}\ }\textbf {\bibinfo {volume} {5}},\ \bibinfo {pages}
  {489} (\bibinfo {year} {2009})}\BibitemShut {NoStop}%
\bibitem [{\citenamefont {Chan}\ \emph {et~al.}(2011)\citenamefont {Chan},
  \citenamefont {Alegre}, \citenamefont {Safavi-Naeini}, \citenamefont {Hill},
  \citenamefont {Krause}, \citenamefont {Gröblacher}, \citenamefont
  {Aspelmeyer},\ and\ \citenamefont {Painter}}]{Chan2011}%
  \BibitemOpen
  \bibfield  {author} {\bibinfo {author} {\bibfnamefont {J.}~\bibnamefont
  {Chan}}, \bibinfo {author} {\bibfnamefont {T.~P.~M.}\ \bibnamefont {Alegre}},
  \bibinfo {author} {\bibfnamefont {A.~H.}\ \bibnamefont {Safavi-Naeini}},
  \bibinfo {author} {\bibfnamefont {J.~T.}\ \bibnamefont {Hill}}, \bibinfo
  {author} {\bibfnamefont {A.}~\bibnamefont {Krause}}, \bibinfo {author}
  {\bibfnamefont {S.}~\bibnamefont {Gröblacher}}, \bibinfo {author}
  {\bibfnamefont {M.}~\bibnamefont {Aspelmeyer}},\ and\ \bibinfo {author}
  {\bibfnamefont {O.}~\bibnamefont {Painter}},\ }\bibfield  {title} {\bibinfo
  {title} {Laser cooling of a nanomechanical oscillator into its quantum ground
  state},\ }\href {https://doi.org/10.1038/nature10461} {\bibfield  {journal}
  {\bibinfo  {journal} {Nature}\ }\textbf {\bibinfo {volume} {478}},\ \bibinfo
  {pages} {89} (\bibinfo {year} {2011})}\BibitemShut {NoStop}%
\bibitem [{\citenamefont {Schliesser}\ \emph {et~al.}(2009)\citenamefont
  {Schliesser}, \citenamefont {Arcizet}, \citenamefont {Rivi{\`{e}}re},
  \citenamefont {Anetsberger},\ and\ \citenamefont
  {Kippenberg}}]{Schliesser2009}%
  \BibitemOpen
  \bibfield  {author} {\bibinfo {author} {\bibfnamefont {A.}~\bibnamefont
  {Schliesser}}, \bibinfo {author} {\bibfnamefont {O.}~\bibnamefont {Arcizet}},
  \bibinfo {author} {\bibfnamefont {R.}~\bibnamefont {Rivi{\`{e}}re}}, \bibinfo
  {author} {\bibfnamefont {G.}~\bibnamefont {Anetsberger}},\ and\ \bibinfo
  {author} {\bibfnamefont {T.~J.}\ \bibnamefont {Kippenberg}},\ }\bibfield
  {title} {\bibinfo {title} {Resolved-sideband cooling and position measurement
  of a micromechanical oscillator close to the heisenberg uncertainty limit},\
  }\href {https://doi.org/10.1038/nphys1304} {\bibfield  {journal} {\bibinfo
  {journal} {Nat. Phys.}\ }\textbf {\bibinfo {volume} {5}},\ \bibinfo {pages}
  {509} (\bibinfo {year} {2009})}\BibitemShut {NoStop}%
\bibitem [{\citenamefont {Riedinger}\ \emph {et~al.}(2016)\citenamefont
  {Riedinger}, \citenamefont {Hong}, \citenamefont {Norte}, \citenamefont
  {Slater}, \citenamefont {Shang}, \citenamefont {Krause}, \citenamefont
  {Anant}, \citenamefont {Aspelmeyer},\ and\ \citenamefont
  {Gröblacher}}]{Riedinger2016}%
  \BibitemOpen
  \bibfield  {author} {\bibinfo {author} {\bibfnamefont {R.}~\bibnamefont
  {Riedinger}}, \bibinfo {author} {\bibfnamefont {S.}~\bibnamefont {Hong}},
  \bibinfo {author} {\bibfnamefont {R.~A.}\ \bibnamefont {Norte}}, \bibinfo
  {author} {\bibfnamefont {J.~A.}\ \bibnamefont {Slater}}, \bibinfo {author}
  {\bibfnamefont {J.}~\bibnamefont {Shang}}, \bibinfo {author} {\bibfnamefont
  {A.~G.}\ \bibnamefont {Krause}}, \bibinfo {author} {\bibfnamefont
  {V.}~\bibnamefont {Anant}}, \bibinfo {author} {\bibfnamefont
  {M.}~\bibnamefont {Aspelmeyer}},\ and\ \bibinfo {author} {\bibfnamefont
  {S.}~\bibnamefont {Gröblacher}},\ }\bibfield  {title} {\bibinfo {title}
  {Non-classical correlations between single photons and phonons from a
  mechanical oscillator},\ }\href {https://doi.org/10.1038/nature16536}
  {\bibfield  {journal} {\bibinfo  {journal} {Nature}\ }\textbf {\bibinfo
  {volume} {530}},\ \bibinfo {pages} {313} (\bibinfo {year}
  {2016})}\BibitemShut {NoStop}%
\bibitem [{\citenamefont {Ding}\ \emph {et~al.}(2011)\citenamefont {Ding},
  \citenamefont {Baker}, \citenamefont {Senellart}, \citenamefont {Lemaitre},
  \citenamefont {Ducci}, \citenamefont {Leo},\ and\ \citenamefont
  {Favero}}]{Ding2011}%
  \BibitemOpen
  \bibfield  {author} {\bibinfo {author} {\bibfnamefont {L.}~\bibnamefont
  {Ding}}, \bibinfo {author} {\bibfnamefont {C.}~\bibnamefont {Baker}},
  \bibinfo {author} {\bibfnamefont {P.}~\bibnamefont {Senellart}}, \bibinfo
  {author} {\bibfnamefont {A.}~\bibnamefont {Lemaitre}}, \bibinfo {author}
  {\bibfnamefont {S.}~\bibnamefont {Ducci}}, \bibinfo {author} {\bibfnamefont
  {G.}~\bibnamefont {Leo}},\ and\ \bibinfo {author} {\bibfnamefont
  {I.}~\bibnamefont {Favero}},\ }\bibfield  {title} {\bibinfo {title}
  {Wavelength-sized {GaAs} optomechanical resonators with gigahertz
  frequency},\ }\href {https://doi.org/10.1063/1.3563711} {\bibfield  {journal}
  {\bibinfo  {journal} {Appl. Phys. Lett.}\ }\textbf {\bibinfo {volume} {98}},\
  \bibinfo {pages} {113108} (\bibinfo {year} {2011})}\BibitemShut {NoStop}%
\bibitem [{\citenamefont {Krause}\ \emph {et~al.}(2015)\citenamefont {Krause},
  \citenamefont {Hill}, \citenamefont {Ludwig}, \citenamefont {Safavi-Naeini},
  \citenamefont {Chan}, \citenamefont {Marquardt},\ and\ \citenamefont
  {Painter}}]{Krause2015}%
  \BibitemOpen
  \bibfield  {author} {\bibinfo {author} {\bibfnamefont {A.~G.}\ \bibnamefont
  {Krause}}, \bibinfo {author} {\bibfnamefont {J.~T.}\ \bibnamefont {Hill}},
  \bibinfo {author} {\bibfnamefont {M.}~\bibnamefont {Ludwig}}, \bibinfo
  {author} {\bibfnamefont {A.~H.}\ \bibnamefont {Safavi-Naeini}}, \bibinfo
  {author} {\bibfnamefont {J.}~\bibnamefont {Chan}}, \bibinfo {author}
  {\bibfnamefont {F.}~\bibnamefont {Marquardt}},\ and\ \bibinfo {author}
  {\bibfnamefont {O.}~\bibnamefont {Painter}},\ }\bibfield  {title} {\bibinfo
  {title} {Nonlinear radiation pressure dynamics in an optomechanical
  crystal},\ }\href {https://doi.org/10.1103/physrevlett.115.233601} {\bibfield
   {journal} {\bibinfo  {journal} {Phys. Rev. Lett.}\ }\textbf {\bibinfo
  {volume} {115}},\ \bibinfo {pages} {233601} (\bibinfo {year}
  {2015})}\BibitemShut {NoStop}%
\end{thebibliography}%

\end{document}